\documentclass[useAMS,usenatbib]{mn2e}

\usepackage{graphicx}

\title[]{AzTEC 1.1 mm images of 16 radio galaxies at 0.5$<$z$<$5.2 and a quasar at z$=$6.3}
\author[A. Humphrey et al.]{A. Humphrey$^{1}$\thanks{E-mail: ajh@inaoep.mx}, M. Zeballos$^{1}$, I. Aretxaga$^{1}$, D.~H. Hughes$^{1}$, M.~S. Yun$^{2}$, R. Cybulski$^{2}$,
\newauthor Grant W. Wilson$^{2}$, J. Austermann$^{3}$, H. Ezawa$^{4}$, R. Kawabe$^{4}$, K. Kohno$^{5,6}$,
\newauthor T. Perera$^{7}$, K. Scott$^{2,8}$, D. S\'anchez-Arguelles$^{1}$, R. Gutermuth$^{9}$\\
$^{1}$Instituto Nacional de Astrof\'{\i}sica, \'Optica y Electr\'onica (INAOE), Aptdo. Postal 51 y 216, 72000 Puebla, Pue., Mexico\\
$^{2}$Department of Astronomy, University of Massachusetts, Amherst, MA 01003, USA\\
$^{3}$Center for Astrophysics and Space Astronomy, University of Colorado, Boulder, CO 80309, USA\\
$^{4}$Nobeyama Radio Observatory, National Astronomical Observatory of Japan, Minamimaki, Minamisaku, Nagano 384-1305, Japan\\
$^{5}$Institute of Astronomy, University of Tokyo, 2-21-1 Osawa, Mitaka, Tokyo 181-0015, Japan\\
$^{6}$Research Center for the Early Universe, University of Tokyo, 7-3-1 Hongo, Bunkyo, Tokyo 113-0033, Japan\\
$^{7}$Department of Physics, Illinois Wesleyan University, Bloomington, IL 61701, USA\\
$^{8}$Department of Physics and Astronomy, University of Pennsylvania, Philadelphia, PA 19104, USA\\
$^{9}$Smith College, Northampton, MA 01063, USA}
\begin{document}

\date{Accepted 2011 July 13. Received 2011 July 11; in original form 2011 February 28}

\pagerange{\pageref{firstpage}--\pageref{lastpage}} \pubyear{2011}

\maketitle

\label{firstpage}

\begin{abstract}
We present 1.1 mm observations for a sample of 16 powerful radio galaxies at 0.5$<$z$<$5.2 and a radio quiet quasar at z=6.3, obtained using the AzTEC bolometer array mounted on the ASTE or the JCMT.  This paper more than doubles the number of high-z radio galaxies imaged at millimetre/sub-millimetre wavelengths.  We detect probable millimetre-wave counterparts for 11 of the active galaxies.  The 6 active galaxies which do not have a probable millimetre counterpart in our images nevertheless have one or more likely associated millimetric source.  Thus, we conclude that powerful (radio-loud) active galaxies at high-z are beacons for finding luminous millimetre/sub-millimetre galaxies at high-z.  The flux densities of our AzTEC {\it counterparts} imply star formation rates ranging from $<$200 to $\sim$1300 $M_{\odot}$ yr$^{-1}$.  In addition, we find that for the radio galaxoes the 1.1 mm flux density is anticorrelated with the largest angular size of the radio source.

We also present new {\it Spitzer} imaging observations of several active galaxies in our sample.  Combining these with archival data, we examine the mid-infrared colours of our sample.  We find that radio galaxies for which we have detected a probable 1.1 mm counterpart have mid-infrared colours consistent with dusty starbursts, and are usually bluer than high-z {\it Spitzer}-selected active galaxies.  In addition, we find arcs of 24 $\mu$m sources extending across $\sim$200-500 kpc, apparently associated with three of the radio galaxies.

\end{abstract}


\begin{figure*}
\includegraphics{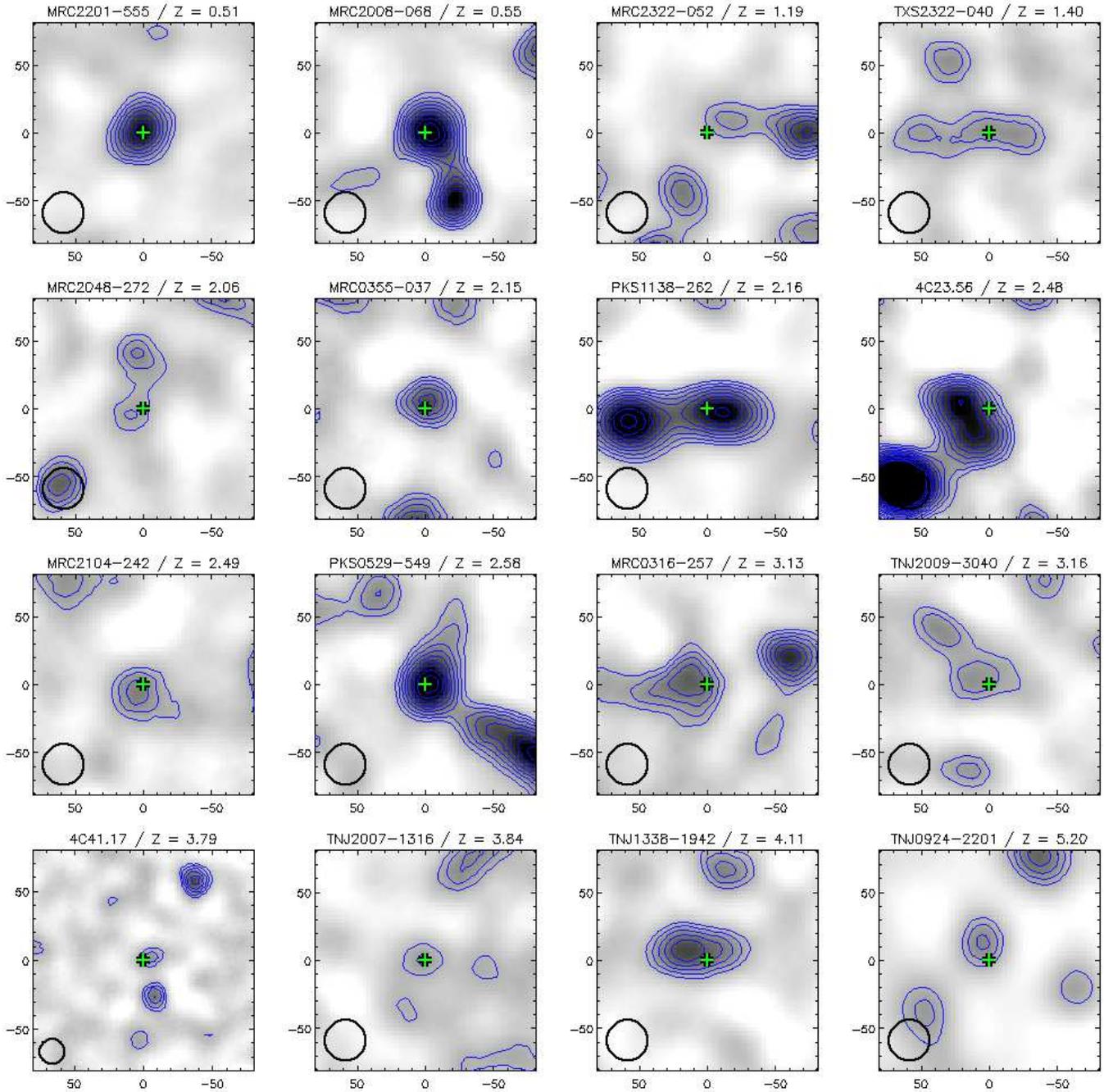}
\vspace{7.1in}
\caption{Postage stamps of 1.1 mm AzTEC signal-to-noise maps for the 16 radio galaxies, which show a 160\arcsec $\times$ 160\arcsec field (or 0.9 Mpc $\times$ 0.9 Mpc to 1.4 Mpc $\times$ 1.4 Mpc in physical units) centred on the radio-optical position (+).  Both axes are labeled in arcsec, with the spatial zero corresponding to the radio-optical position of the radio galaxy.  The galaxies are ordered with z increasing left to right, top to bottom.  Contours start at 2$\sigma$ and increase linearly by 1$\sigma$.  The FWHM of the beam is shown in the lower left corner of each map.  Note that the map of 4C+41.17 has a smaller beam FWHM than the rest of the sample, because the observations were made using the 15 m JCMT (beam FWHM=18\arcsec), rather than the 10 m ASTE (beam FWHM=28\arcsec).  Probable 1.1 mm counterparts for 10 of the radio galaxies are detected at $\ge$3$\sigma$ significance.}
\end{figure*}

\begin{figure}
\includegraphics{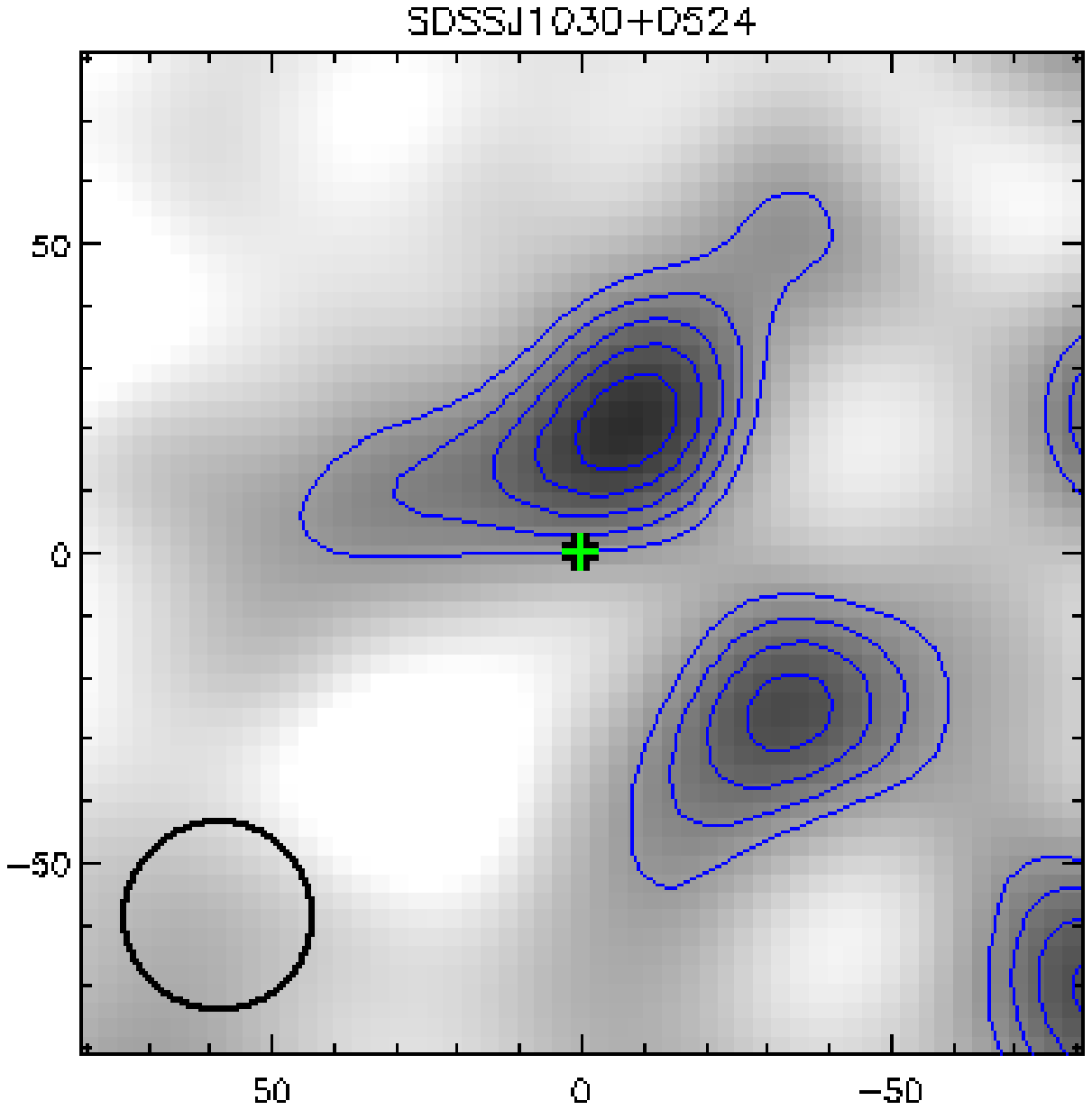}
\vspace{1.5in}
\caption{Postage stamp of the 1.1 mm AzTEC signal-to-noise map for the z=6.3 radio quiet quasar SDSS J1030+0524.  The 160\arcsec $\times$ 160\arcsec field (or 0.9 Mpc $\times$ 0.9 Mpc in physical units) is centred on the optical position, which is marked by a cross.  Contours start at 2$\sigma$ and increase linearly by 1$\sigma$.  The FWHM of the beam is shown in the lower left corner.  We do not detect the 1.1 mm counterpart of this galaxy.}
\end{figure}

\section{Introduction}
Powerful, high-z radio galaxies (z$\ge$0.5: HzRGs) continue to play a key role in cosmological investigations.  Their high luminosities at radio and optical wavelengths make them useful as beacons for finding massive elliptical galaxies and their progenitors out to high redshifts (e.g. Dunlop et al. 1996; R\"ottgering et al. 1997; McLure et al. 1999), and provide unique opportunities to study their host galaxy and environment (e.g. Venemans et al. 2007).  

Many HzRGs are embedded within haloes of ionized gas which are extended on spatial scales of tens of kpc, and which emit luminous emission lines from various species (e.g. McCarthy, Spinrad \& van Breugel 1995; Reuland et al. 2003a; Villar-Mart{\'{\i}}n et al. 2003).  Spatially resolved kinematic studies of these haloes suggest that the cold gas comprising many of these haloes is in infall towards the centre of the host galaxy (Humphrey et al. 2007), while in others it is being strongly disturbed by the radio jets (e.g. Best, R\"ottgering \& Longair 2000; Nesvadba et al. 2006).  In most cases, ratios between emission lines imply that the radiation field of the AGN is the dominant ionization mechanism for the haloes, although shocks may make a fractional contribution in some cases (e.g. Vernet et al. 2001; Humphrey et al. 2008).  Modelling of the flux ratio Ly$\alpha$/HeII$\lambda$1640 measured from one-dimensional spectra has revealed an excess of Ly$\alpha$ emission above the predictions of AGN photoionization models (Villar-Mart{\'{\i}}n et al. 2007), which suggests the presence of star forming regions in the giant gaseous haloes (see also Hatch et al. 2009).  

Millimetre emission provides an alternative means by which the star formation history of HzRGs can be probed.  Several tens of HzRGs have now been detected at millimetre wavelengths, with the detection rate rising from $\sim$15 per cent at z$<$2.5 to $\ga$75 per cent at z$>$2.5 (Archibald et al. 2001; Reuland et al. 2003).  Millimetric observations are sensitive to cold dust that is re-radiating emission received from young stars: implied rates of star formation range from a few hundred to $\sim$1500 $M_{\odot}$ yr$^{-1}$ (Dunlop et al. 1994; Archibald et al. 2001; Reuland et al. 2004).  These millimetre/sub-millimetre detected radio galaxies are not typical of the wider population of millimetre/sub-millimetre selected galaxies (MMGs hereinafter): unlike the vast majority of MMGs, the mm emission from radio galaxies is often extended on spatial scales of $\ga$100 kpc (e.g. Stevens et al. 2003), can show morphological complexity (Greve et al. 2007), and is not always spatially coincident with the radio-optical nucleus of the AGN and host galaxy (Ivison et al. 2008).  Several HzRGs also appear to have MMGs as companions, or are associated with an overdensity of MMGs (Ivison et al. 2000; Stevens et al. 2003; see also Priddey, Ivison \& Isaak 2008 and Stevens et al. 2010).  

In this paper we present images and measure 1.1 mm flux densities for a sample of 16 powerful radio galaxies at 0.5$<$z$<$5.2 and a radio-quiet quasar at z=6.3, observed with the AzTEC bolometer array instrument (Wilson et al. 2008) mounted on the Atacama Submillimetre Telescope Experiment (ASTE: Ezawa et al. 2004) or the James Clerk Maxwell Telescope (JCMT), as part of the AzTEC/ASTE Cluster Environment Survey (ACES: Zeballos et al., in preparation).  Throughout this paper we adopt a flat Universe with $H_0$=71 km s$^{-1}$ Mpc$^{-1}$, $\Omega_M$=0.27 and $\Omega_{\Lambda}$=0.73.  In this cosmology 1\arcsec corresponds to 5.7-8.6 kpc for 0.5$<$z$<$6.3.

\begin{table*}
\centering
\caption{Basic properties of the sample of radio galaxies.  Columns are as follow.  (1) Name of the active galaxy.  (2) Active galaxy redshift.  (3) Rest-frame wavelength, in units of $\mu$m, corresponding to 1.1 mm in the observer frame.  (4) and (5) are the RA and Dec of the active galaxy.  (6) The wavelength regime from which the position of the active galaxy was determined, where R = radio, I = mid-IR (3.6 and 4.5 $\mu$m), O = optical (R-band) and K = near-IR (K-band).  (7) 1.1 mm flux density of the millimetre counterpart to the active galaxy.  Errors in $S_{1.1}$ are 1$\sigma$, and upper limits are 3$\sigma$.  (8) The observed offset of the mm counterpart relative to the position of the active galaxy.  In the absence of a probable counterpart we show in parentheses the offset to the closest mm detection.  (9) P-statistic of the probable mm counterpart.  Again, in the absence of a probable counterpart we show in parentheses the P-statistic of the nearest mm detection.  (10) Log of the far-IR luminosity of mm counterparts.  For non-detections, we give 3$\sigma$ upper limits.  (11) The star formation rate implied by the far-IR luminosity.  Included at the bottom of the table is the radio-quiet quasar SDSS J1030+0524.} 
\begin{tabular}{lllllllllll}
\hline
Source name   & z    & $\lambda_{rest}$ & RA (J2000) & Dec (J2000) & ID & $S_{1.1}$ & r & P & Log $(L_{FIR}/L_{\odot})$ & SFR \\
              &      & ($\mu$m)         &(hh:mm:ss.ss)&(dd:mm:ss.s)&    & (mJy)   &(\arcsec)& &            & ($M_{\odot}$ yr$^{-1}$) \\
(1)           & (2)  & (3)              & (4)        & (5)         & (6)& (7)       & (8)   &(9)& (10)     & (11)  \\    
\hline
MRC 2201-555  & 0.51 & 730& 22:05:04.83& -55:17:44.0& I& 6.1$\pm$0.8 & 2.5    & 8$\times$10$^{-6}$ & 12.86    & 720    \\
MRC 2008-068  & 0.55 & 710& 20:11:14.22& -06:44:03.6& R& 8.6$\pm$0.9 & 3.5    & 4$\times$10$^{-6}$ & 13.04    & 1100   \\
MRC 2322-052  & 1.19 & 500& 23:25:19.62& -04:57:36.6& I& $<$2.1      & (20.7) & (0.049)            & $<$12.60 & $<$400 \\
TXS 2322-040  & 1.51 & 440& 23:25:10.23& -03:44:46.7& R& 2.3$\pm$0.6 & 2.3    & 0.0006             & 12.68    & 480    \\
MRC 2048-272  & 2.06 & 360& 20:51:03.49& -27:03:03.7& I& 2.3$\pm$0.8 & 11.4   & 0.02               & 12.68    & 480    \\
MRC 0355-037  & 2.15 & 350& 03:57:48.06& -03:34:09.5& I& 3.5$\pm$0.7 & 6.6    & 0.002              & 13.02    & 1040   \\
PKS 1138-262  & 2.16 & 350& 11:40:48.35& -26:29:08.6& R& $<$3.6      & (14.3) & (0.007)            & $<$12.85 & $<$710 \\
4C +23.56     & 2.48 & 320& 21:07:14.82& +23:31:45.1& R& $<$1.5      & (21.4) & (0.005)            & $<$12.46 & $<$290 \\
MRC 2104-242  & 2.49 & 320& 21:06:58.27& -24:05:09.1& R& 3.7$\pm$0.9 & 6.9    & 0.002              & 12.87    & 740    \\
PKS 0529-549  & 2.58 & 310& 05:30:25.43& -54:54:23.3& I& 6.4$\pm$0.7 & 3.7    & 2$\times$10$^{-5}$ & 13.10    & 1270   \\
MRC 0316-257  & 3.13 & 270& 03:18:12.14& -25:35:10.2& I& $<$2.1      & (17.3) & (0.009)            & $<$12.55 & $<$350 \\
TN J2009-3040 & 3.16 & 260& 20:09:48.08& -30:40:07.4& K& 3.3$\pm$0.9 & 12.9   & 0.005              & 12.79    & 610    \\
4C +41.17     & 3.79 & 230& 06:50:52.35& +41:30:31.4& O& 3.5$\pm$1.0 & 6.6    & 0.001              & 12.78    & 610    \\
TN J2007-1316 & 3.83 & 230& 20:07:53.22& -13:16:43.4& R& 2.8$\pm$0.9 & 4.0    & 0.002              & 12.68    & 480    \\
TN J1338-1942 & 4.10 & 220& 13:38:26.23& -19:42:33.6& R& $<$5.0      & (15.7) & (0.0008)           & $<$12.90 & $<$790 \\
TN J0924-2201 & 5.19 & 180& 09:24:19.91& -22:01:41.5& I& 3.6$\pm$0.9 & 13.9   & 0.006              & 12.74    & 560    \\
\hline
SDSS J1030+0524&6.28 & 150 & 10:30:27.10& +05:24:55.0& I& $<$1.4     & (22.7) & (0.02)             & $<$12.30 & $<$200 \\
\hline
\end{tabular}
\end{table*}

\section{Data}

\subsection{Sample}
For this investigation we have selected 16 powerful radio galaxies at 0.5$<$z$<$5.2 from compendia of high-z radio galaxies (e.g. Miley \& De Breuck 2008).  This sample is designed to cover a large range in z, to afford investigations into possible cosmic evolution of the properties of powerful radio galaxies and their environments.  The lower limit to z was set at 0.5 in order to minimise the possible contribution to the observer-frame 1.1 mm emission from the Sunyaev-Zel'dovich effect (SZE), though it is not impossible that the SZE still makes a significant (i.e., a few mJy) contribution to the 1.1 mm flux densities of radio galaxies at 0.5$<$z$\la$1 (e.g. High et al. 2010; Menanteau et al. 2010; Foley et al. 2011).  However, one of our galaxies in that redshift range (MRC 2008-068: z=0.55) has been imaged in X-rays (XMM-Newton observation ID 0502510501), but shows no evidence for the extended X-ray emitting gas needed to produce a significant SZE signal.  Since we have no evidence for any SZE signal in the 0.5$<$z$\la$1 radio galaxies, we are unable to make a realistic correction for possible SZE contamination.  

Our sample also covers a substantial range in the (projected) size of the radio source, with diameters ranging between 0.2 kpc and 390 kpc, in order to study the possible relationship between the growth or containment of the radio source and the millimetric properties of the galaxy.  The sample purposefully contains some radio galaxies that are known to reside in clusters or proto-clusters (e.g. PKS 1138-262: Pentericci et al. 1997), some radio galaxies which do not appear to reside in overdense environments (e.g. MRC 2048-272: Venemans et al. 2007), as well as radio galaxies for which cluster/overdensity analyses have yet to be undertaken.  

In the process of selecting our sample, no strong preference was given to any other observational properties of the galaxies, but several strong biases necessarily exist due to processes involved in the initial identification of the radio galaxies.  Firstly, the galaxies are all near the top of the radio luminosity function for radio galaxies, and the majority of them were selected from radio catalogues on the basis of their ultra-steep radio spectra ($S_v \propto v^{\alpha}$ where $\alpha \la$-1: e.g. Chambers et al. 1996).  In addition, these galaxies all have luminous UV-optical continuum emission, as required for the identification of the optical counterpart.  They also all emit luminous UV-optical emission lines from spatially extended regions of cold gas, photoionized by the active nucleus: such conditions are necessary for obtaining a spectroscopic redshift.  In summary, our sample is a subset of high-z radio galaxies which have very powerful, steep-spectrum radio sources, together with luminous UV-optical continuum and line emission.  

It is important to note that roughly 30 per cent of high-z radio galaxy candidates do not show bright optical/UV emission lines (e.g. Miley \& De Breuck 2008), meaning that samples selected in the above way may not be completely representative of high-z radio galaxies.  Interestingly, it has been suggested that these 'no-line' radio galaxies are very dusty, based on millimetre/sub-millimetre detections thereof (Reuland et al. 2003b; Reuland 2005).  

Our sample also includes the z=6.28 radio quiet quasar SDSS J103027.10+052455.0 (Fan et al. 2001).  Throughout this paper, we abbreviate this source name to J1030+0524.  When the observations were being prepared, this source held the record for the active galaxy with the highest known z.  It is currently unclear whether this quasar resides in a rich environment (Stiavelli et al. 2005; Willott et al. 2005, Priddey, Ivison \& Isaak 2008; Kim et al. 2009).

\subsection{AzTEC 1.1 mm observations}
In the pilot of this programme, the field of 4C+41.17 was mapped utilising unchopped raster scans, using the AzTEC bolometer array (Wilson et al. 2008) with the James Clerk Maxwell Telescope (JCMT) during December 2005.  The rest of our sample were observed during May to October 2007 or July to December 2008, with AzTEC at the Atacama Submillimetre Telescope Experiment (ASTE), using a Lissajous pattern centred on the targeted active galaxy.  The observations were made under excellent weather conditions, with effective atmospheric opacities for each field falling within the range 0.03$\le$$\tau$$\le$0.06.  Integration times varied between 16 and 35 h per field.  The resulting maps cover areas ranging from 170 to 300 arcmin$^2$ for a 50 per cent coverage cut.  

The data reduction procedure is given in Scott et al. (2008) with modifications discussed by Scott et al. (2010) and Downes et al. (2011).  The basic steps are (i) to clean the raw time-stream data of spikes due to cosmic rays and instrumental glitches; (ii) to then clean using principal component analysis; (iii) to calibrate and bin the bolometer signals to produce a map for each individual observation, with a pixel size of 2\arcsec$\times$2\arcsec for 4C+41.17 or 3\arcsec$\times$3\arcsec for the rest of the sample; and (iv) to co-add maps and apply a Wiener filter.  The observations, the data reduction and the source detection/extraction are described in greater detail by Zeballos et al. (in preparation).  
 
\subsection{\emph{Spitzer} Imaging}
The \emph{Spitzer} (Werner et al. 2004) observations of the active galaxies come from both archival data and observations from PID 50610 (PI: M.~S. Yun).  Fifteen of the galaxies were observed with the Multiband Imaging Photometer for Spitzer (MIPS: Rieke et al. 2004) at 24 $\mu$m, and fourteen were observed with the Infrared Array Camera (IRAC: Fazio et al. 2004) at 3.6, 4.5, 5.8, and 8.0 $\mu$m.  Our new observations (PID 50610) consist of 5 MIPS maps (MRC 2201-555; MRC 2008-068; MRC 0355-037; TN J2009-3040; TN J1338-1942) and 3 IRAC maps (MRC 2201-555; MRC 2322-052; MRC 0355-037) of HzRG fields. The new IRAC maps are composed of a 4x3 grid of pointing positions, and at each position there are 15 dithered frames of 100 sec exposure time each. The new MIPS maps are comprised of a 6x6 grid of pointing positions, and with each position mapped twice witn an individual 30 sec exposure time.

From the Spitzer Science Center's basic calibrated data (BCD) frames we build mosaics for each field using a customized IDL package (Gutermuth et al. 2008).  A number of common bright source artifacts (banding, jailbars, muxbleed, and pulldown effects) are corrected in the individual BCD frames as a first step. Then the code performs a rejection of transients and accounts for distortions introduced by rotation and sub-pixel offsets in building the mosaics. One additional calibration step, performed only on the MIPS observations, is a self-calibration to remove systematic artifacts (like bright and dark latents) from the maps. For each MIPS astronomical observation request (AOR), a flat-field map is made by taking the median of the background at each BCD pixel. Then the median background map is normalized, and all BCD images for that AOR are divided by the normalized flat field map. The final calibrated, corrected mosaics for IRAC and MIPS were then resampled to a scale of 0.865$\arcsec pix^{-1}$ and 1.80$\arcsec pix^{-1}$, respectively. 

Photometric measurements were performed on the mosaics using the IDL routine APER with an aperture diameter of 3.8$\arcsec$. We apply aperture corrections in the IRAC bands of [1.40, 1.38, 1.55, 1.70], and for MIPS we use a correction of 2.00. The corrections are determined by selecting isolated sources, having no neighboring detection at 5$\sigma$ within 10$\arcsec$ identified by Source Extractor (Bertin \& Arnouts 1996), and taking the ratio of the fluxes of those isolated sources with apertures of 12.4 and 3.8$\arcsec$ diameter. The differences between our aperture corrections and those listed in the IRAC Instrument Handbook are as one would expected for our 3.8$\arcsec$ diameter photometry aperture. For each map the median of these flux ratios is taken, and then the overall aperture correction is chosen as the mean of the set of corrections derived for each band. Small field-to-field variations in the median aperture corrections introduce some systematic uncertainties in our photometry, and so the standard deviation of the aperture corrections derived in our maps is added in quadrature to the statistical photometric errors for the \emph{Spitzer} bands.  In the event that we do not detect an active galaxy, we give a 3$\sigma$ upper limit.  

Ten of the radio galaxies in our sample have \emph{Spitzer} photometry reported in the literature (Villar-Mart{\'{\i}}n et al. 2006; Seymour et al. 2007).  For all but one of these (PKS 0529-549), our photometry is consistent with the previously published measurements.  In the case of PKS 0529-549, we measure the 24 $\mu$m flux to be significantly lower than that reported by Seymour et al. (2007): we obtain 632$\pm$27 $\mu$Jy compared to the 942$\pm$71 $\mu$Jy of Seymour et al. (2007).  The higher flux density measured by Seymour et al. is probably due to the presence of a second 24 $\mu$m source located $\sim$6\arcsec SW of the radio galaxy, which would have contributed significantly to the light measured in their relatively large 13\arcsec diameter aperture, but significantly less so in our 3.8\arcsec aperture.  

\section{Results}
Figs. 1 and 2 show 160\arcsec$\times$160\arcsec~ 'postage stamps' cut out from the full-size signal-to-noise maps, with contours starting at 2$\sigma$ and increasing linearly by 1$\sigma$.  Bars indicate the orientation of the radio source, where applicable.  Most of the active galaxies in our sample have been observed at a variety of wavelengths and spatial resolutions, meaning that there are various different ways by which one can define the position of the galaxy.  For this study, we will adopt the coordinates that are most likely to mark the position of the galactic nucleus.  Our preferred position is that of the radio core, since it corresponds to the active nucleus itself.  For active galaxies for which the radio core has not been detected, or which are radio quiet, our second choice is to use the centroid of the {\it SPITZER} IRAC emission, averaged across the 3.6 and 4.5 $\mu$m bands.  For type 2 (i.e., narrow line) objects this emission is expected to trace the evolved stellar population, which is likely to be more dynamically relaxed than young stellar populations which, if present, could dominate at shorter optical-IR and longer (i.e. MIPS) wavelengths.  For type 1 (broad line) objects, this emission is likely to be dominated by the active nucleus.  Finally, in the absence of both a radio core detection and IRAC data, we adopt the position of the optical or near-IR emission.  The postage stamp images are centred on our fiducial 'radio-optical' position, the coordinates of which are listed in Table 1.  

\subsection{Association between millimetre sources and radio galaxies}
In this paper we adopt a source detection threshold such that the peak S/N is $\ge$3.0 and also that the spatial FWHM is consistent with, or larger than, the full width at half maximum (FWHM) of the filtered point spread function (i.e., FWHM$\ge$34\arcsec).  We have attempted to identify sources associated with our active galaxies as follows.  First, we have calculated the Poisson probability that a detected mm source is merely a chance association, rather than being associated with the active galaxy.  This is given by 

\begin{equation}
P=1-e^{-\pi r^2 N}
\end{equation}

\noindent where $r$ is the angular distance of the mm source from the optical/IR/radio position of the active galaxy, and $N$ is the surface density of mm sources in blank fields that have 1.1 mm flux densities greater than or equal to that of the detected mm source (Downes et al. 1986).  For $N$ we adopt the number counts from the AzTEC/SHADES survey (Austermann et al. 2010).  We consider a mm source to be associated with the active galaxy when the P-statistic is less then 0.05, i.e., the null hypothesis of chance association is significant at the $<$5 per cent level.  All 17 of the active galaxies in our sample have at least one associated 1.1 mm source.  
 
However, we must apply an additional criterion in order to determine whether any of the associated mm sources are consistent with being at the position of the active galaxy.  In other words, are the optical/IR/radio and mm positions consistent, taking into account the observational errors in the position of the mm source?  The 1$\sigma$ positional error of an observed mm source is 

\begin{equation}
\sigma_{pos} \simeq \sqrt{\Bigg(\frac{0.6~ FWHM}{S/N_{d}}\Bigg)^2+\sigma_{pnt}^2}
\end{equation}

\noindent where FWHM is that of the filtered beam, S/N$_{d}$ is the {\it de-boosted} signal to noise ratio of the millimetre counterpart, and $\sigma_{pnt}$ is the 1$\sigma$ uncertainty in the pointing of the map (adapted from Ivison et al. 2007).  We conservatively assume that $\sigma_{pnt}$$\sim$1\arcsec (Scott et al. 2010).  As an illustration, a source with S/N$_{d}$=3.0 would have a positional error $\sigma_{pos}\simeq$6.9\arcsec.  Note that equation (2) does not take into account any systematic absolute offset in the map, i.e., residuals after pointing corrections.  

We classify an AzTEC source as the 1.1 mm counterpart to the radio galaxy or quasar when P$<$0.05 and when the spatial offset from the radio-optical position $r$ is less than 2.7$\sigma_{pos}$\footnote{The positional error distribution is non-Gaussian.  2.7$\sigma_{pos}$ corresponds to a 95 per cent cumulative probability.}.  There are 11 AzTEC sources which satisfy our criteria for classification as 1.1 mm counterparts to the radio galaxies or quasar.  Notice that 6 active galaxies have both P$\le$0.05 and r$>$2.7$\sigma_{pos}$: this means they have a statistically significant association with a millimetre galaxy which is not likely to be the active galaxy host.  The results from this analysis are shown in Table 1.  

Finally, while the presence of other (bright) sources within $\sim$30-60\arcsec can potentially skew the centroid of a millimetre source due to the negative side-lobes of the telescope beam, our tests show this effect is negligible for our entire sample: all such centroid shifts are $\le$1\arcsec and do not affect our counterpart vs. non-counterpart classification.

\subsection{Consistency with previous  millimetre/sub-millimetre measurements}
A number of the galaxies in our sample have been imaged previously at millimetre or sub-millimetre wavelengths.  These are 4C+41.17 (Ivison et al. 2000; Stevens et al. 2003; Greve et al. 2007), PKS 1138-262 (Stevens et al. 2003), TN J1338-1942 (De Breuck et al. 2004) and SDSS J1030+0524 (Priddey, Ivison \& Isaak 2007).  Taking into account the (usually) lower spatial resolution of our new observations, we find good agreement with the images and photometry from the literature.  In two cases (PKS 1138-262 and TN J1338-1942), several discrete sources detected by SCUBA or MAMBO are blended together in our images, due to their relatively lower spatial resolution.  

In addition, eight of our sample have been targeted with SCUBA photometry (4C+23.56, 4C+41.17: Archibald et al. 2001; PKS 1138-262, MRC 0316-257, TN J2007-1316, TN J1338-1942, TN J0924-2201: Reuland et al. 2004; SDSS 1030+0524: Priddey et al. 2003).  For seven of these galaxies, we find good agreement between our flux density measurements and the SCUBA photometry.  (We use a scaling factor of 2.5 between 850$\mu$m and 1.1 mm, assuming $T_{dust}$=40 K and $\beta$=1.5: see $\S$3.5.)  

TN J0924-2201 is the exception.  Reuland et al. (2004) derived a 3$\sigma$ limit of $\le$3.2 mJy at 850$\mu$m, which implies a 1.1 mm flux density of $\la$1.3 mJy, whereas we measure a 1.1 mm flux density of 3.6$\pm$0.9 mJy.  The reason for this discrepancy is not clear.  It may suggest that the 1.1 mm source we identify with the radio galaxy is not the true `counterpart', and instead is a companion MMG offset from the position of the radio galaxy host (see $\S$6).  Maps with higher spatial resolution and S/N will be needed to address this issue.

\subsection{Stacking}
Next we derive average values of $S_{1.1}$ for various subsets of our sample.  To do this we stack the AzTEC images at the position of the active galaxy, allowing us to include information from the individual observations wherein the active galaxy counterpart was not detected at $\ge$3$\sigma$ significance.  First, we spatially registered the images.  In the case of 4C +41.17, we resampled the image to a scale of 3\arcsec pixel$^{-1}$.  Next, we assigned to each image a weighting proportional to $\sigma^{-2}$.  We then stacked all the images in our sample (Fig. 3, top left).  From this stacked image, we measure an average $S_{1.1}$=3.4$\pm$0.2 mJy at the position of the active galaxy.  We repeat this using only the radio galaxy images, and obtain an average $S_{1.1}$=3.2$\pm$0.2 mJy (Fig. 3, top right).  Repeating for the 6 active galaxies for which we did not detect a millimetre counterpart, we measure an average flux density $S_{1.1}$=2.0$\pm$0.3 mJy (Fig. 3, middle left).  Finally, for the 5 undetected radio galaxies we stack to obtain an average flux density $S_{1.1}$=2.4$\pm$0.3 mJy (Fig. 3, middle right).  

We have also investigated whether the millimetre emission is more extended along the radio axis of the radio galaxies in our sample.  This is motivated in part by the fact that the ultraviolet and optical emission from powerful radio galaxies is usually aligned with the radio source (e.g. McCarthy et al. 1987; Chambers, Miley \& van Breugel 1987), and the apparent alignment between the radio source and the large scale distribution of MMGs found by Stevens et al. (2003) in a sample of 7 radio galaxies at 2.2$\le$z$\le$4.3.  To this end, the AzTEC images were registered, rotated such that the radio axis runs horizontally, and subsequently stacked (Fig. 3, bottom left).  Along the radio axis in the stacked image, we measure FWHM=39.3$\pm$0.6\arcsec, compared to FWHM=40.8$\pm$0.6\arcsec measured orthogonally to the radio axis.  Thus, we conclude that there is no significant trend for the millimetre emission to be more elongated along the radio axis than it is along the orthogonal axis, at least at the spatial resolution of our observations.  

\begin{figure}
\includegraphics{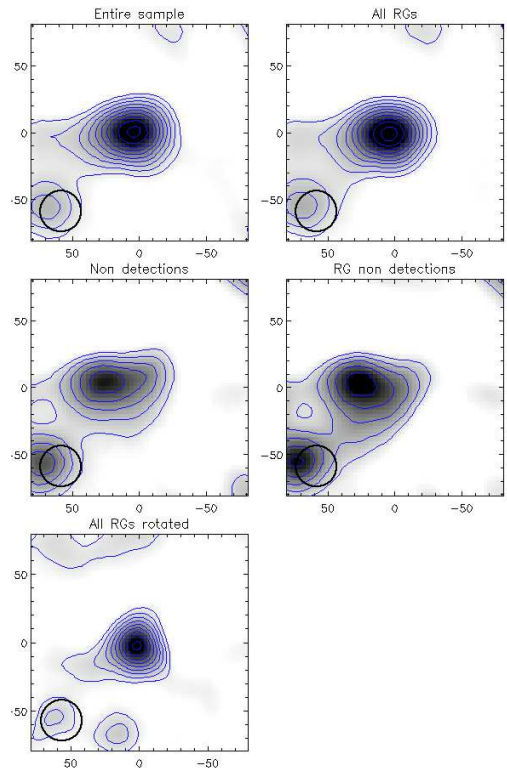}
\vspace{4.0in}
\caption{Postage stamps of the stacked AzTEC images.  The images were shifted such that the radio-optical position of the active galaxy is at the centre, and then co-averaged with weightings proportional to $\sigma^{-2}$.  Top left: all 17 active galaxies in our sample.  Top right: all 16 radio galaxies.  Middle left: all non-detections.  Middle right: non-detected radio galaxies.  Bottom left: radio galaxies stacked with their radio source axes running horizontally.  Contours begin at 2$\sigma$, and increase linearly by 2$\sigma$.}
\end{figure}

\subsection{Synchrotron emission}
Before using the our 1.1 mm flux densities to estimate the infrared luminosity and the implied rate of star formation, we must first consider the impact of synchrotron emission from the radio source at 1.1 mm.  Indeed, previous studies have shown that in some high-z radio galaxies, the synchrotron emission has sufficiently high luminosity and/or a sufficiently flat SED at high frequencies to be detectable in the mm-wave regime (e.g. Archibald et al. 2001; Vieira et al. 2010).  

To quantify the possible contribution at 1.1 mm from synchrotron emission $S_{ext}$, we have extrapolated a power law of the form $S_v\propto$$v^{\alpha_{rad}}$ from the two highest frequency photometric data points of the observed radio spectra separated in $v$ by a factor of $\ge$1.2.  When more than one measurement is available at 'similar' frequencies, which we define as when the ratio between frequencies is $\le$1.2, we have conservatively used the measurement that would result in the highest value of $S_{ext}$ (e.g. MRC 0355-037).  In Fig 4. we show the radio-IR spectral energy distributions of the radio galaxies in our sample, including our extrapolations to 1.1 mm of the synchrotron emission.  

In the first instance, we have used radio photometric measurements that represent flux densities integrated across the entire spatial extent of the radio source, as these data are readily available for all of the radio galaxies in our sample.  However, extrapolating the radio lobe SED to the mm regime is less than ideal.  With the possible exceptions being the two smallest radio galaxies (MRC 2008-068 and TXS 2322-040), this photometry is dominated by the extended radio lobes (see e.g. Carilli et al. 1997; Pentericci et al. 2000).  Lobe spectra usually steepen towards higher frequencies (e.g. Muxlow, Pelletier \& Roland 1988; Murgia et al. 1999); therefore, in most cases this extrapolation is expected to over-predict the contribution from the extended lobes, that is, $S_{ext}$ is expected to represent an extreme upper limit to the contamination due synchrotron emission from the lobes.  Table 2 gives $S_{ext}$ alongside $S_{1.1}$, $\alpha_{rad}$ and the radio wavelengths used for the extrapolations.  The table also gives references to the photometric data points plotted in Fig. 4.  

All 11 of the radio galaxies which we have detected at 1.1 mm have $S_{1.1}$ measurements exceeding $S_{ext}$.  This means that synchrotron emission from the radio lobes clearly cannot be responsible for all of the 1.1 mm emission from these radio galaxies.  Although for 6 of those 11 radio galaxies $S_{ext}$ is greater than the 1$\sigma$ uncertainty on $S_{1.1}$, we refrain from drawing the naive conclusion that $S_{1.1}$ in some cases may be significantly enhanced by synchrotron emission, because our extrapolation of $S_{ext}$ is based on a very favourable set of assumptions, and the results thereof represent rather extreme upper limits.  Indeed, in one particular case (MRC 0316-257) $S_{ext}$ is actually a factor of $\sim$2 {\it higher} than the 3$\sigma$ lower limit to $S_{1.1}$.  

Moving into the millimetre regime from the radio, the flux density of the lobes dimishes more rapidly than that of the core (e.g., Carilli et al. 1997), so that at millimetre wavelengths the core typically dominates the spatially integrated synchrotron flux density (see, e.g., Figure 1 of Nesvadba et al. 2009; Downes et al. 1996).  Thus, we must also consider whether the radio core might contribute significantly to the 1.1 mm emission.  The radio core has been detected in only 4 of our radio galaxy sample (Chambers et al. 1996; Carilli et al. 1997; Pentericci et al. 2000, 2001), and for these galaxies we have extrapolated the radio core SED to 1.1 mm using a powerlaw.  One of these galaxies was detected at 1.1 mm (MRC 2104-242), and the value of $S_{ext}$ for its core is several orders of magnitude below our $S_{1.1}$ measurement (and well below the $S_{ext}$ extrapolated from the radio lobes).

In summary, the radio lobes are unable to provide sufficient synchrotron flux at 1.1 mm to be responsible for the flux densities we measure at that wavelength, despite our use of a very favourable set of assumptions.  Furthermore, it seems likely that any enhancement by synchrotron emission is smaller than the 1$\sigma$ uncertainty in our flux density measurements, i.e., insignificant, in essentially all cases.  Thus, we prefer not to apply any correction for synchrotron contamination to our measured values of $S_{1.1}$, because the extrapolation of the synchrotron flux density is so uncertain.  Finally, we comment that the issue of possible synchrotron contamination would be resolved with observations at longer millimetric wavelengths (e.g. 2-3 mm).

\begin{figure*}
\includegraphics{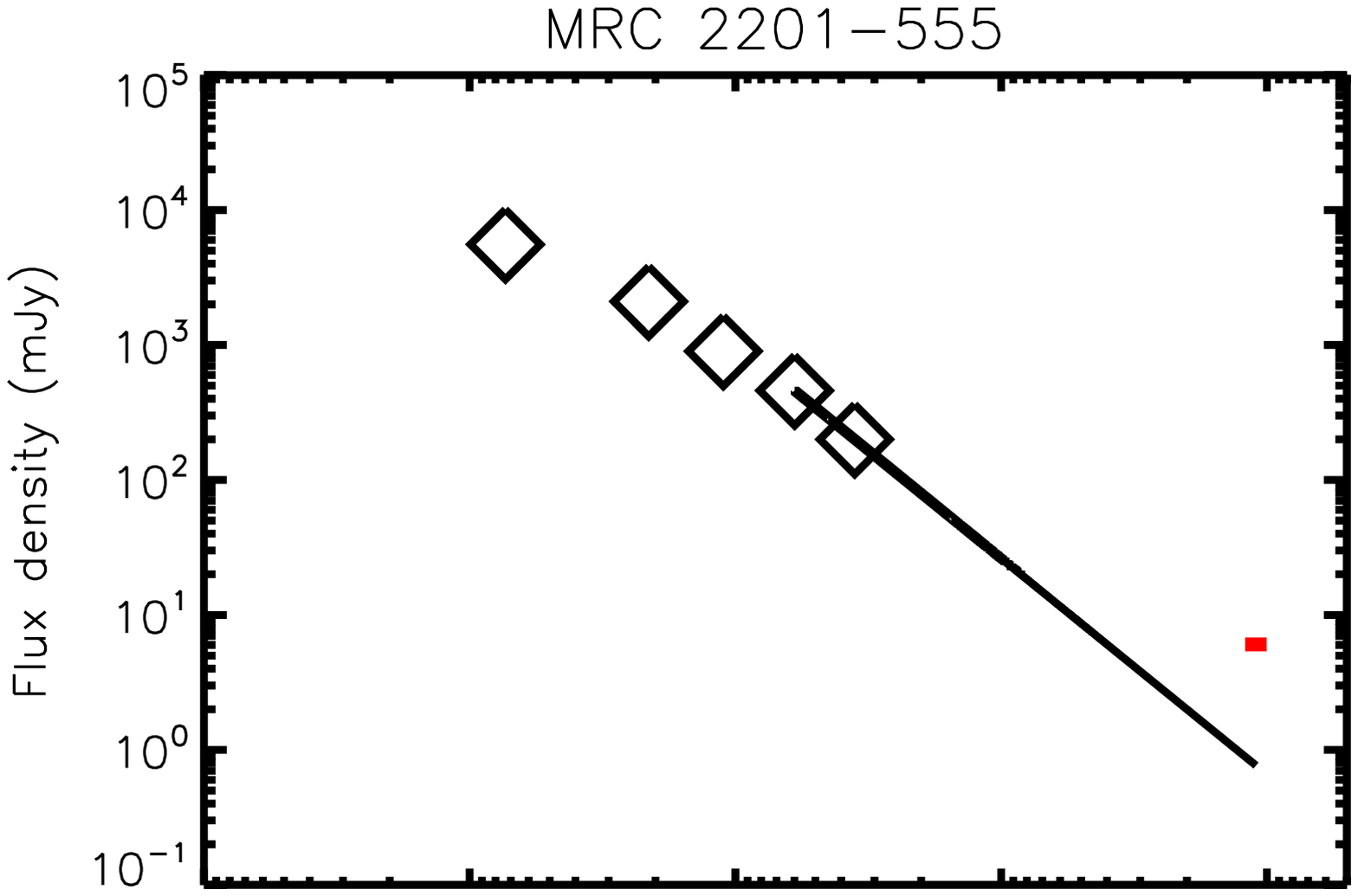}
\includegraphics{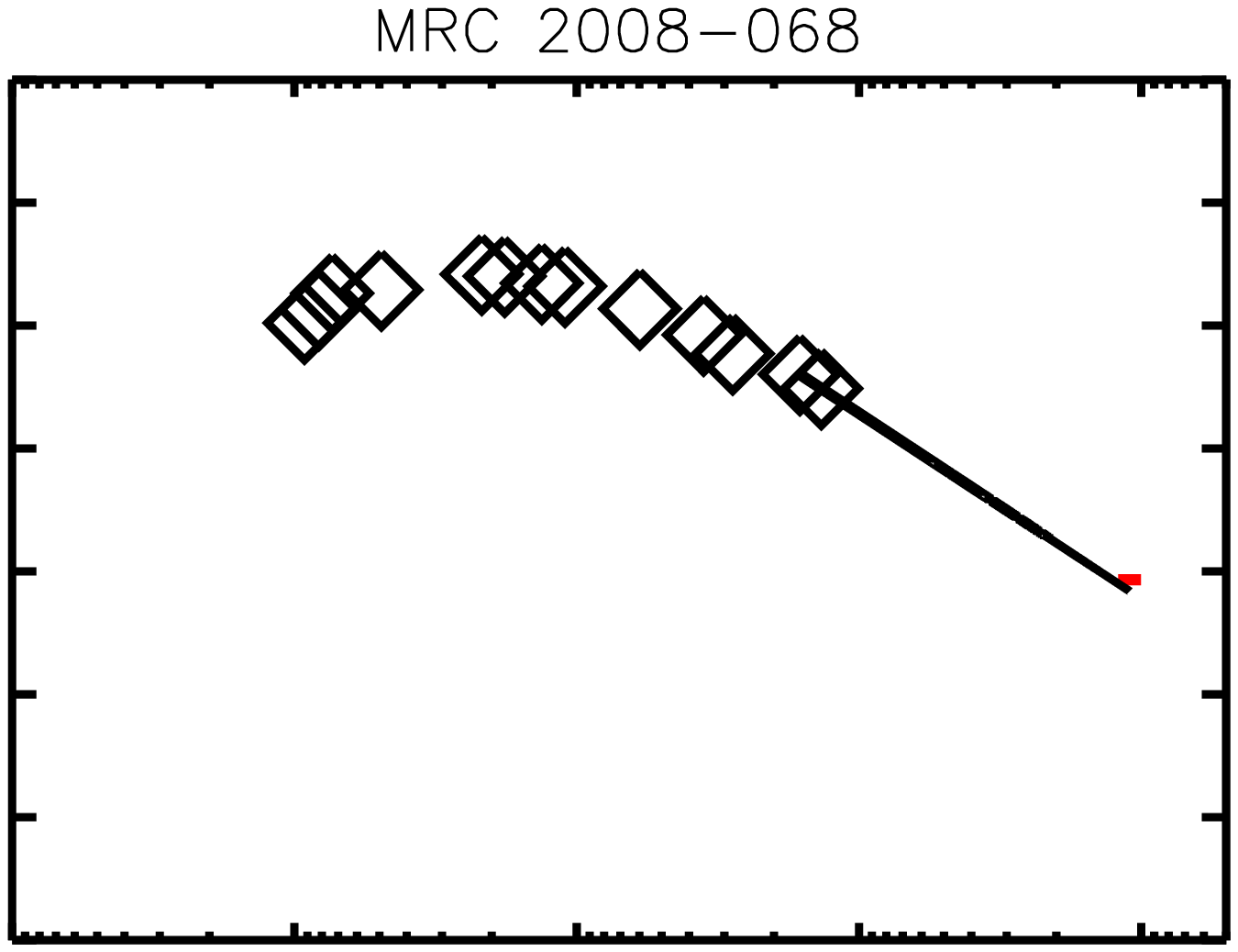}
\includegraphics{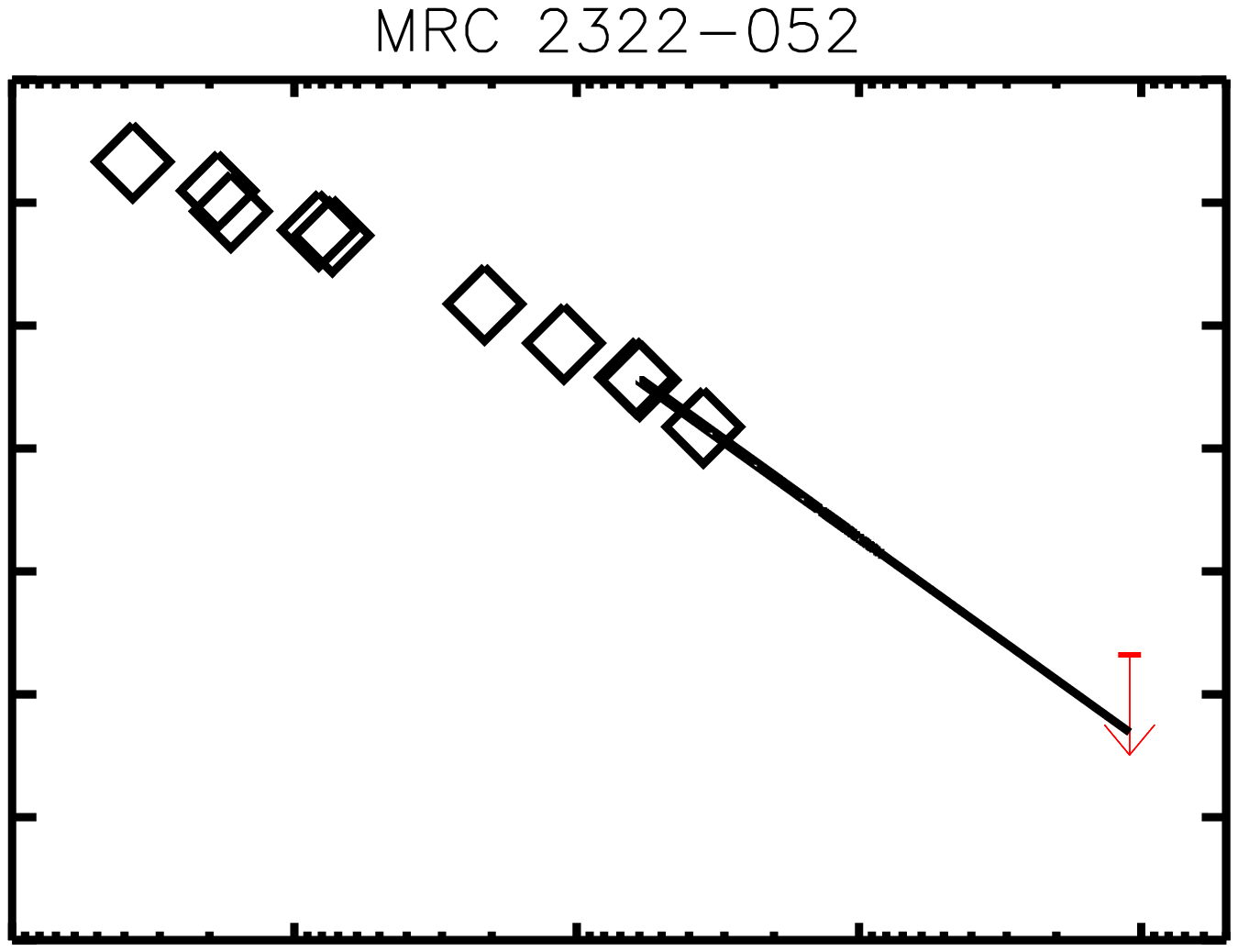}
\includegraphics{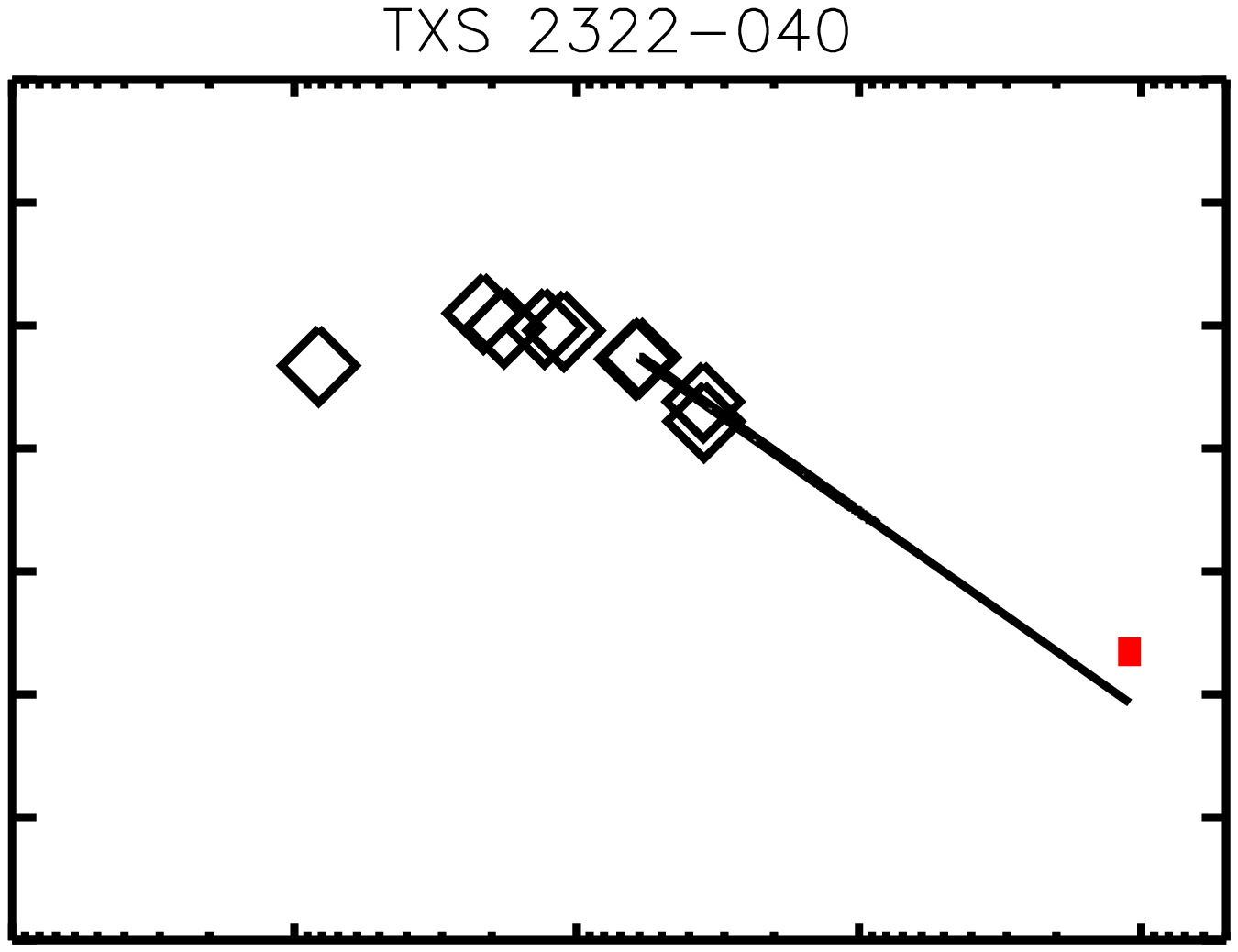}

\includegraphics{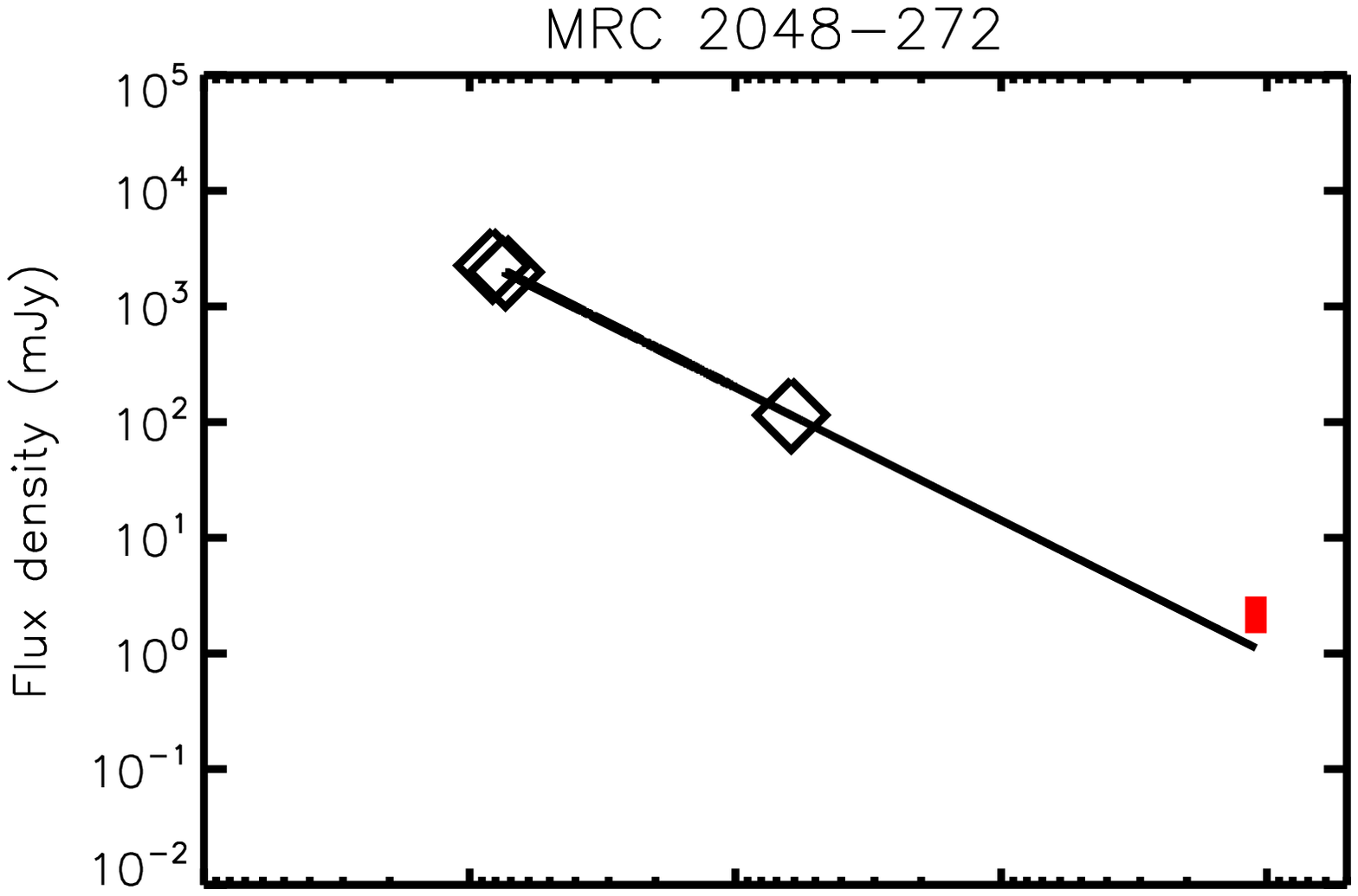}
\includegraphics{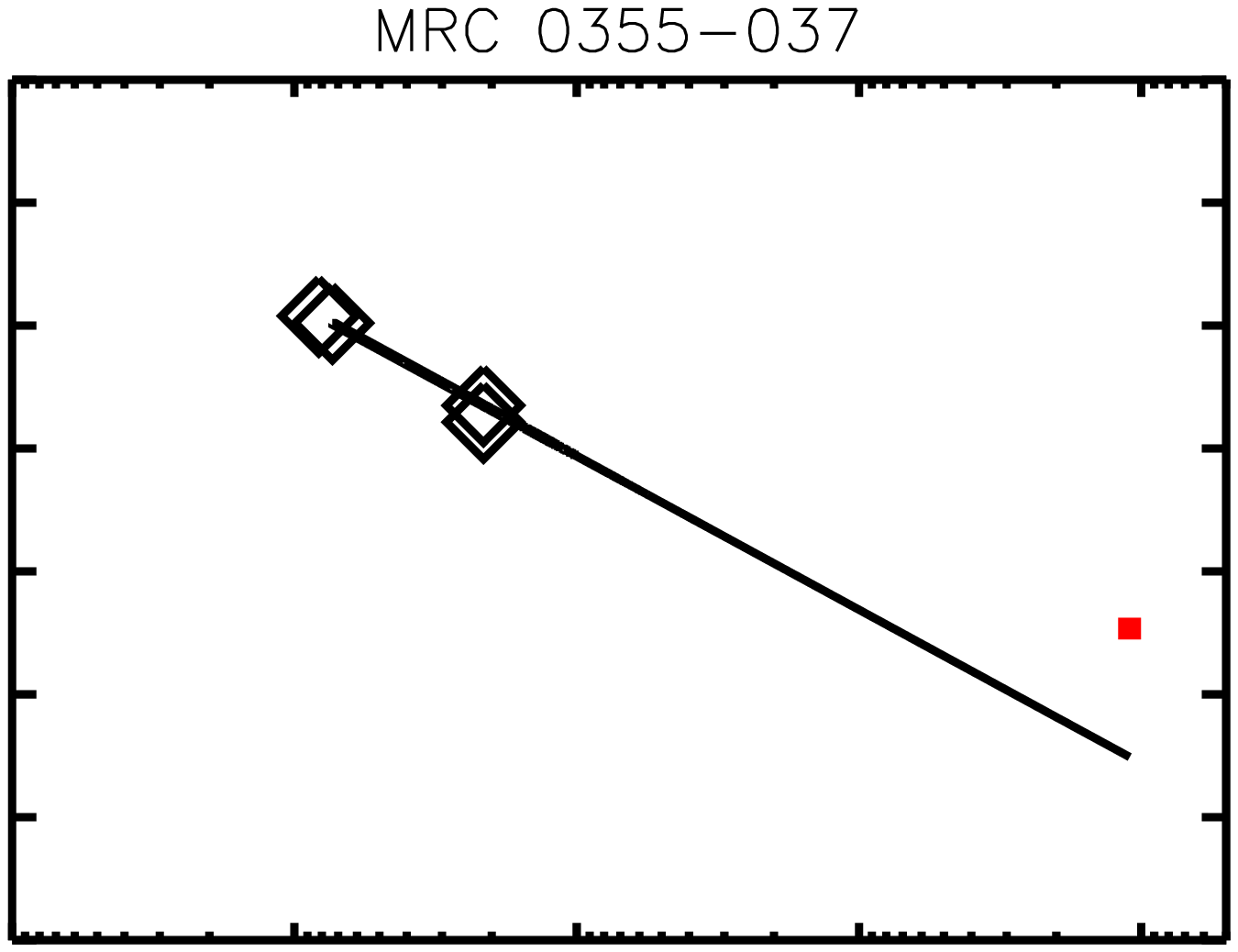}
\includegraphics{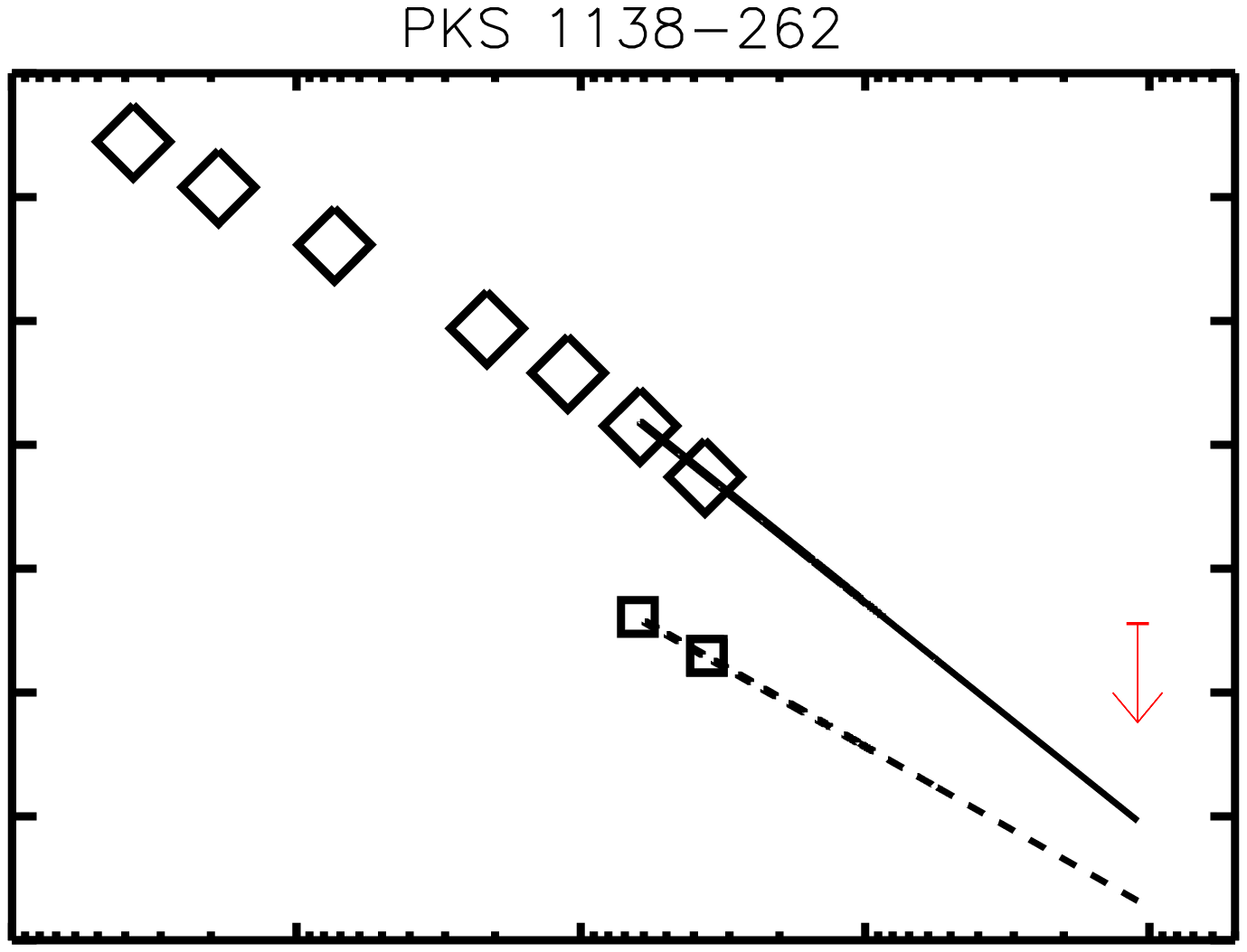}
\includegraphics{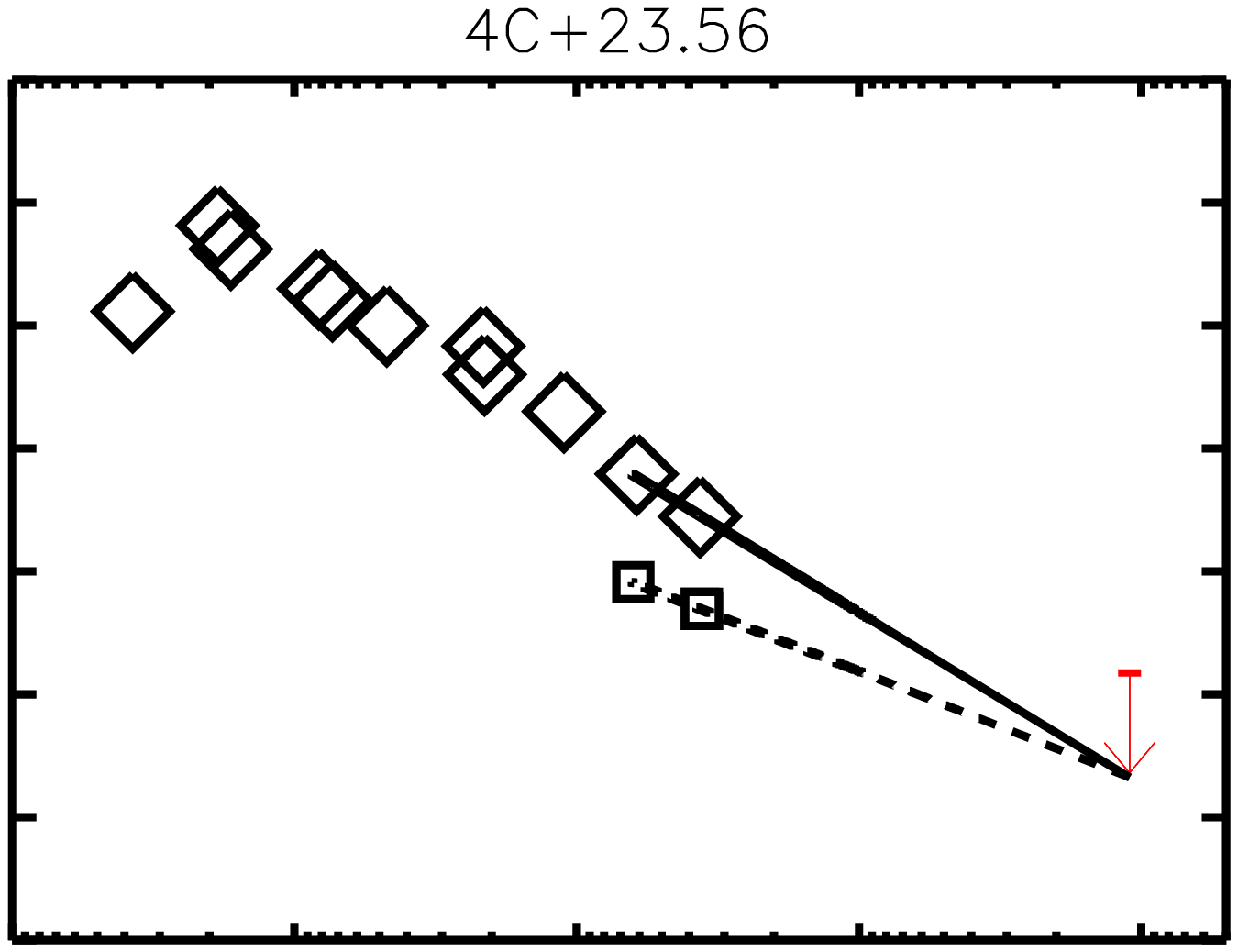}

\includegraphics{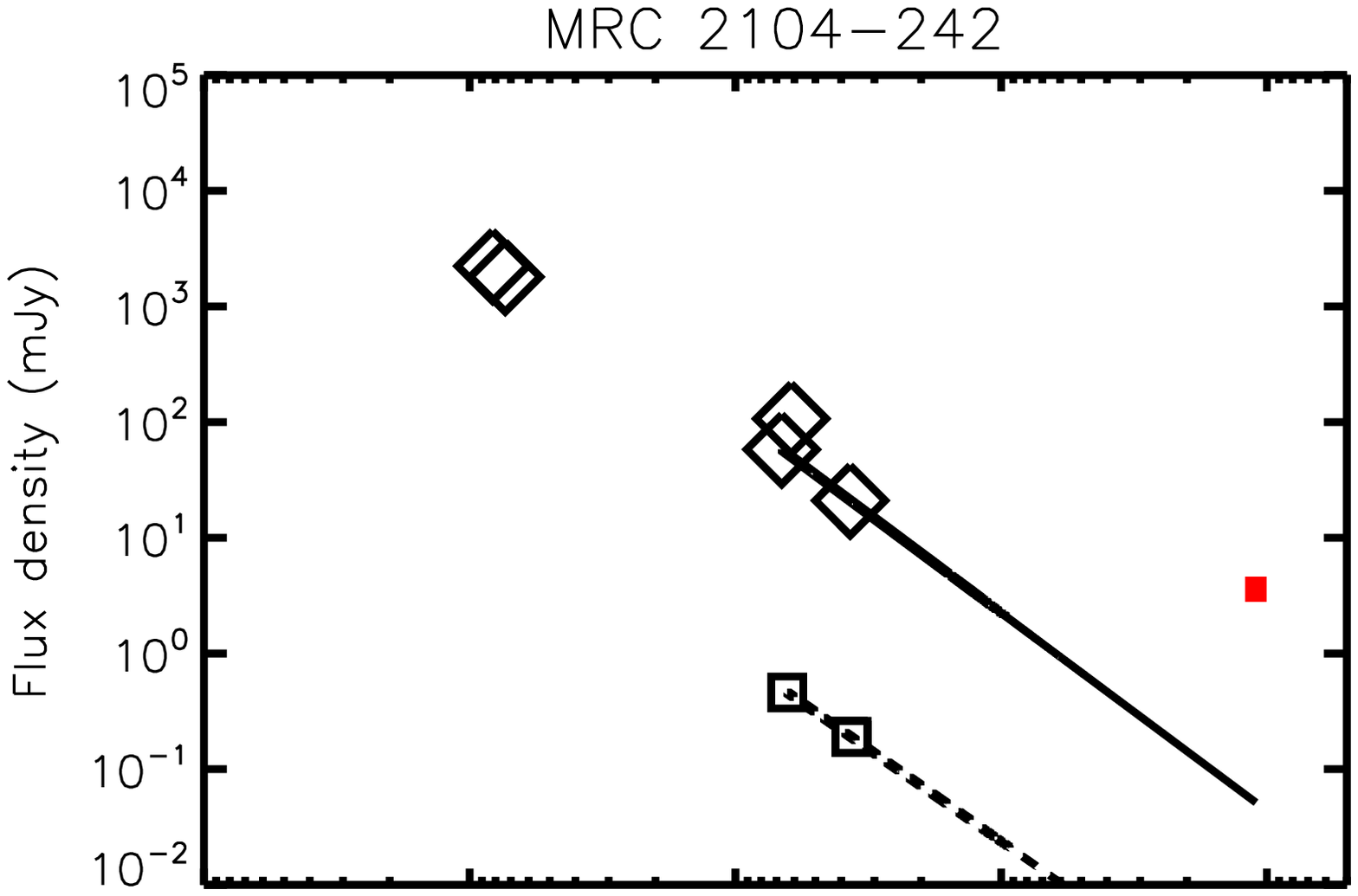}
\includegraphics{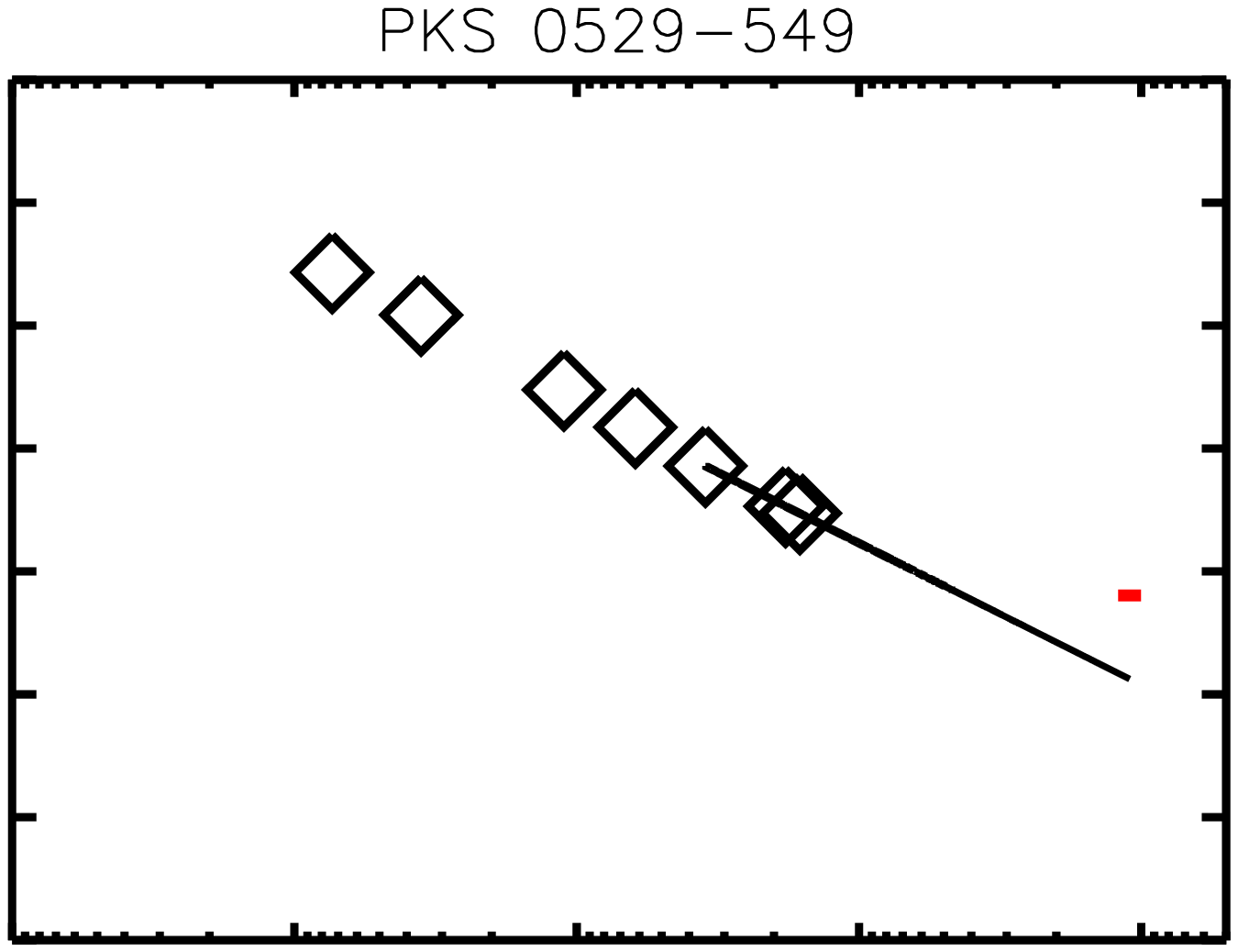}
\includegraphics{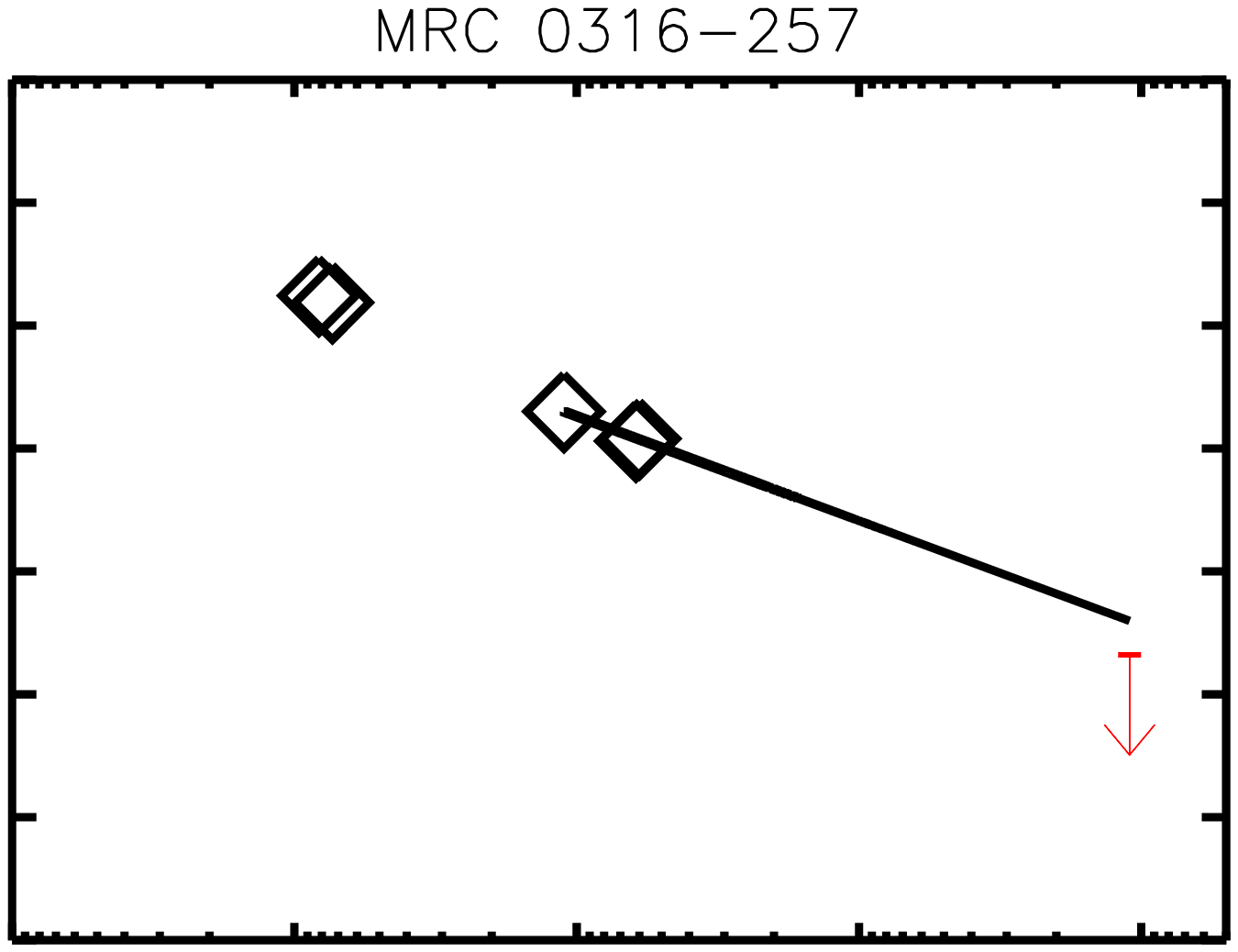}
\includegraphics{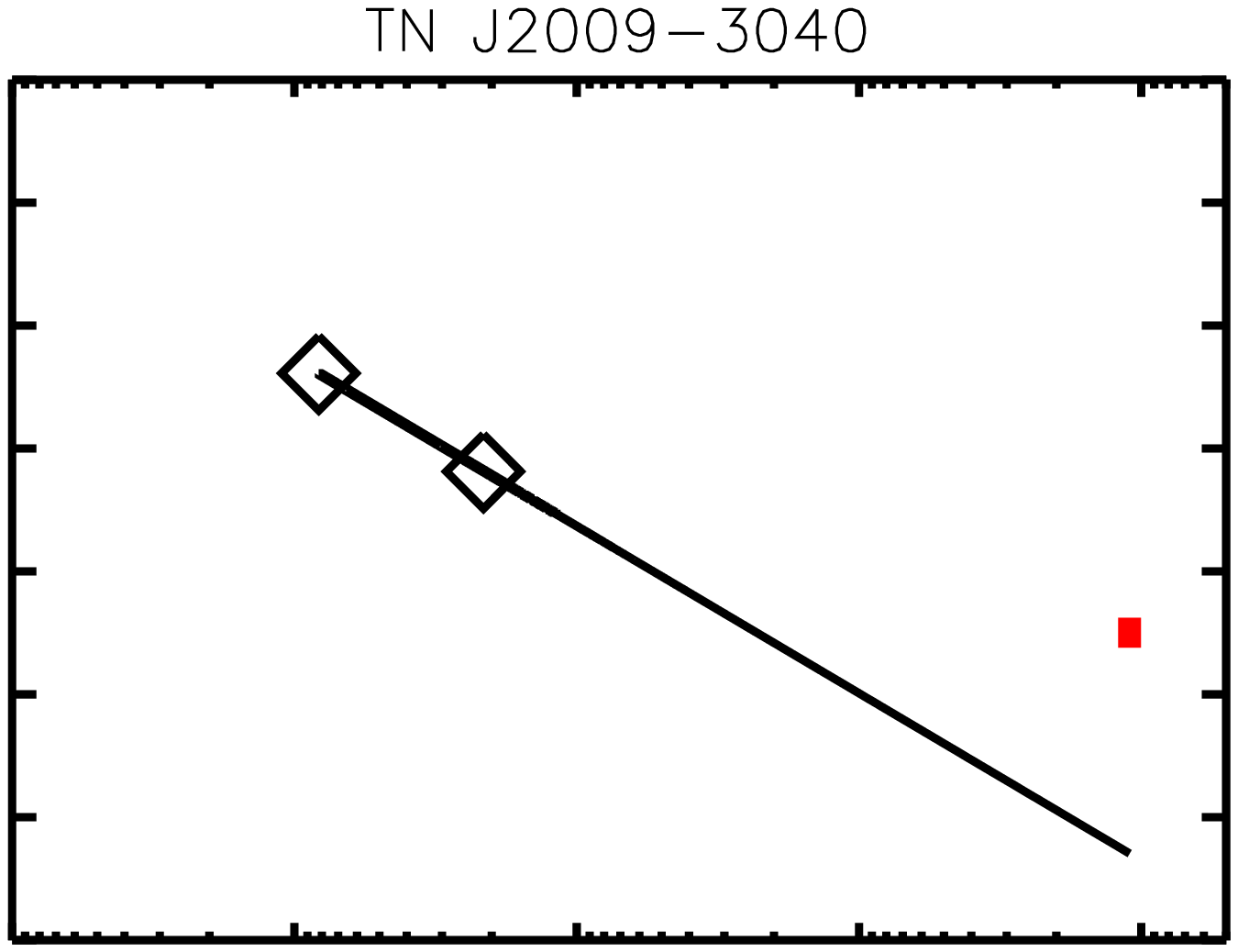}

\includegraphics{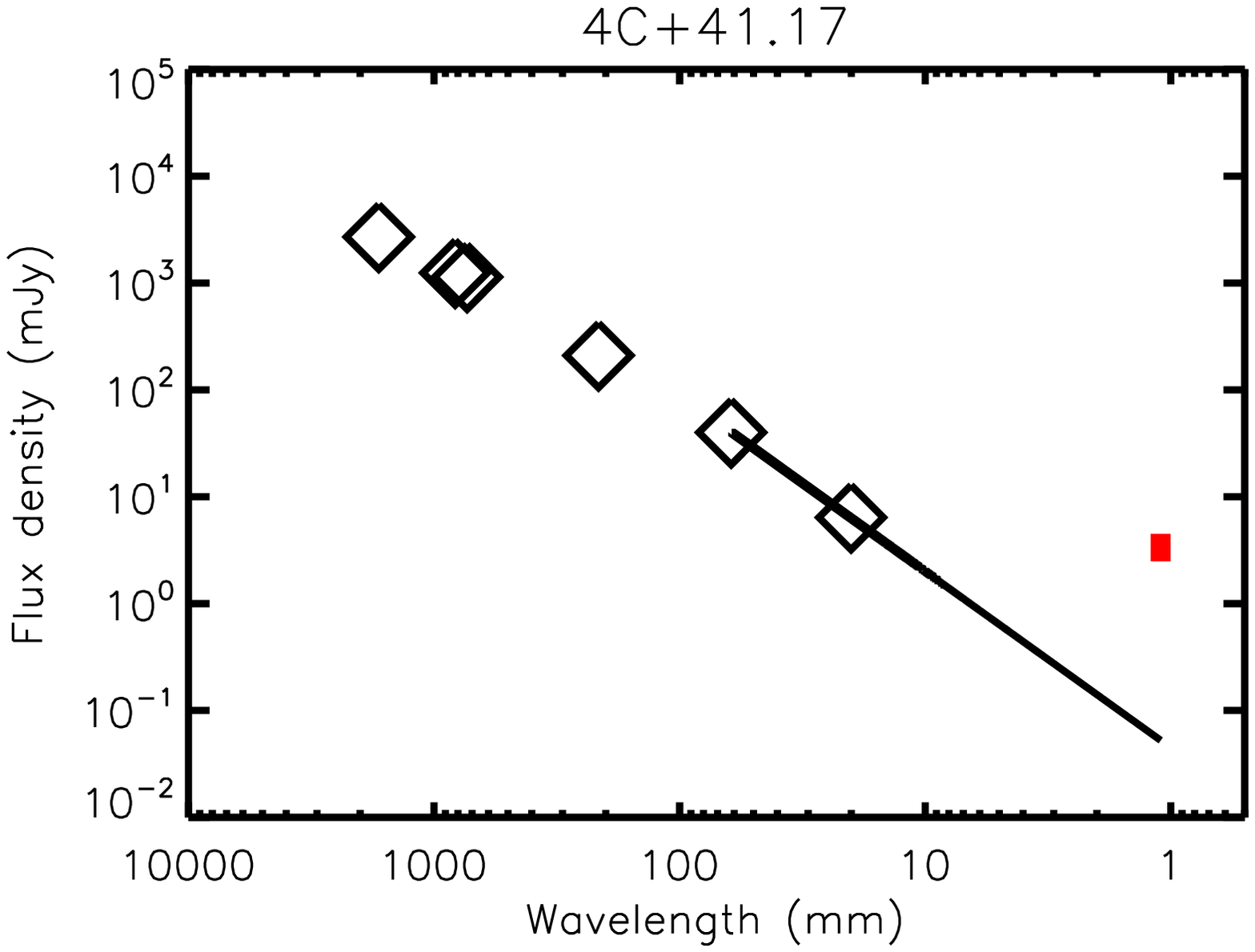}
\includegraphics{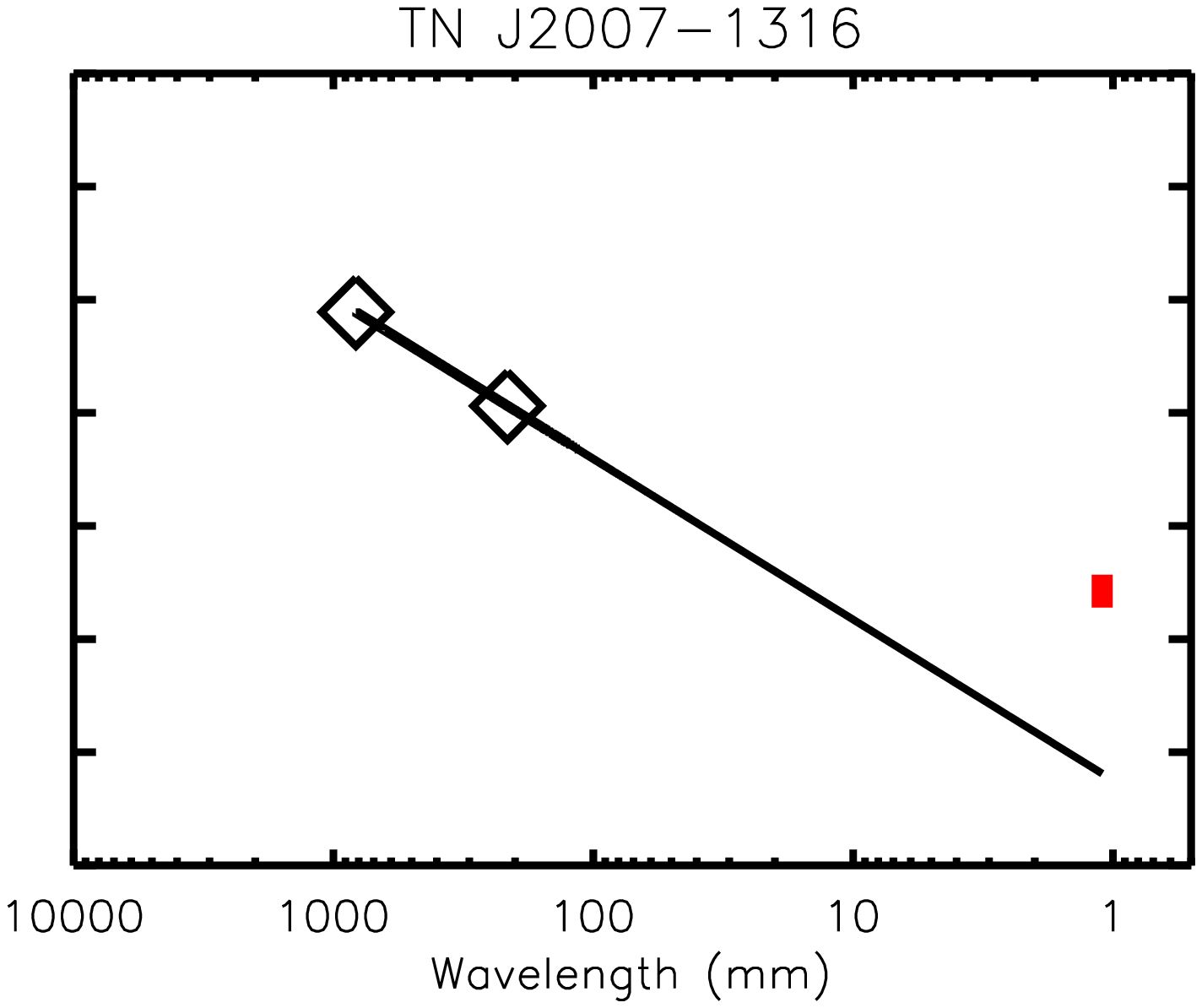}
\includegraphics{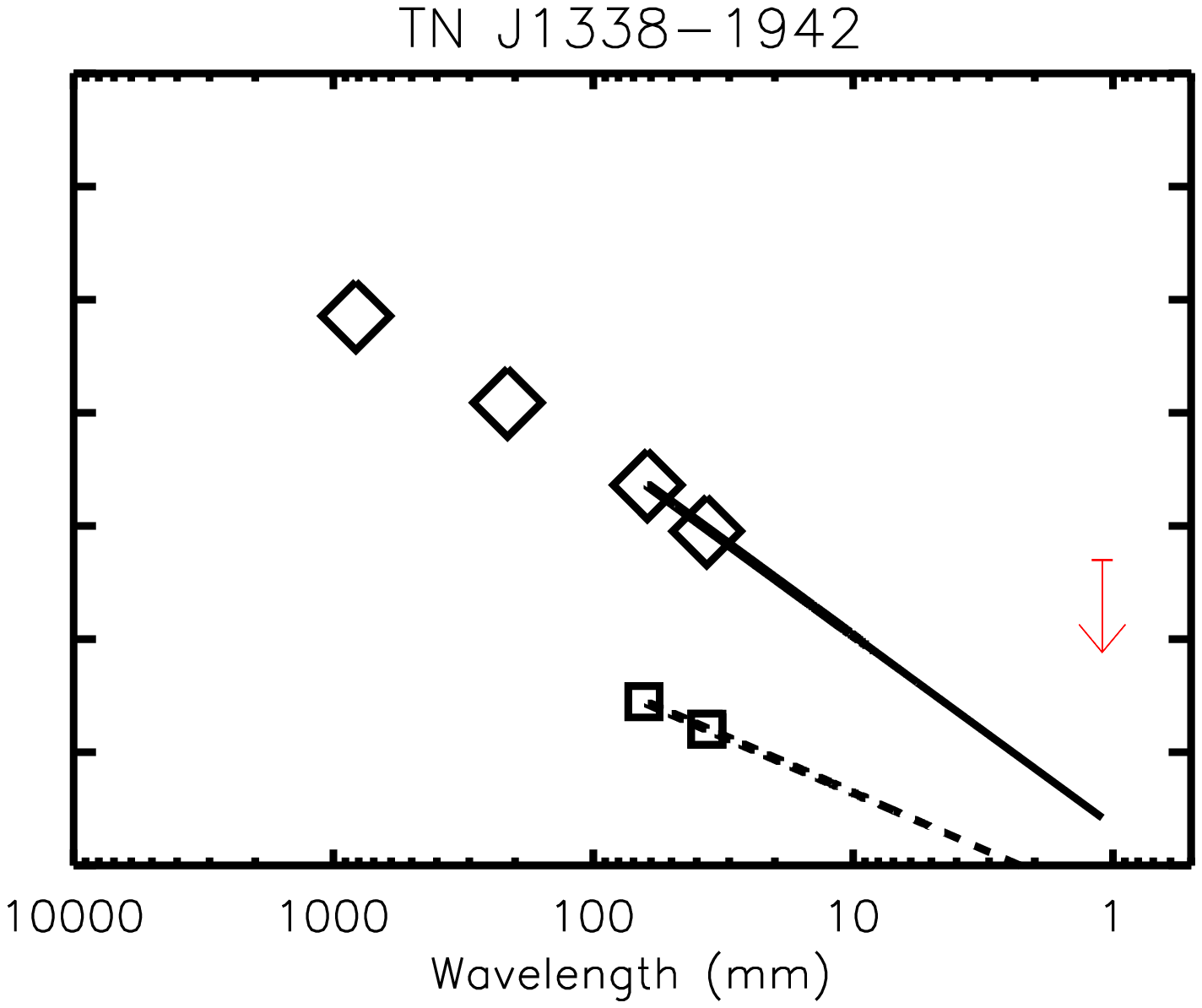}
\includegraphics{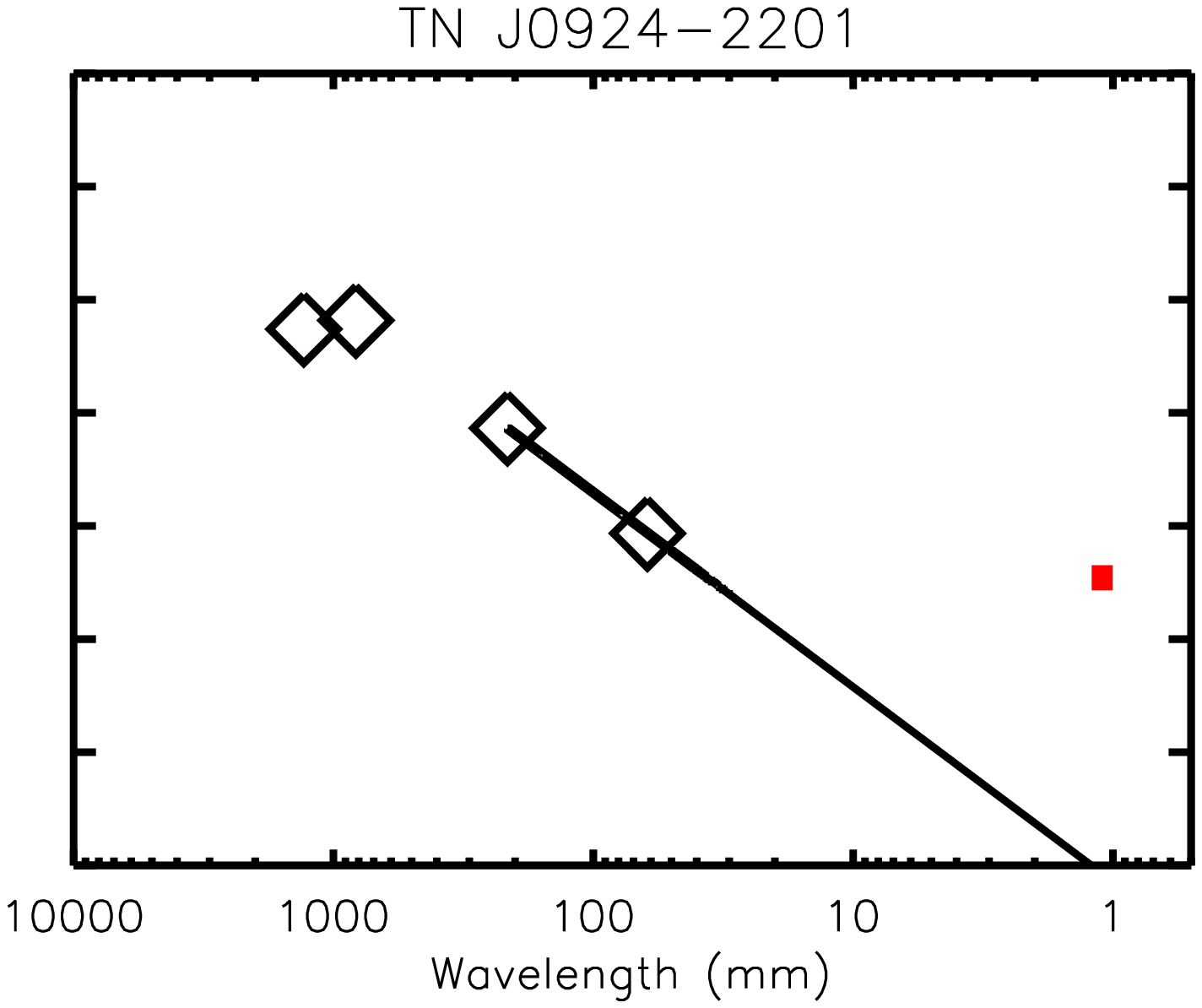}

\vspace{5.5in}
\caption{The observer-frame radio spectral energy distributions of the radio galaxies in our sample and our extrapolation therefrom of extreme upper limits on the 1.1 mm flux density due to synchrotron emission.  The red vertical bars show our AzTEC 1.1 mm flux density measurement, with the vertical length indicating the range of values permitted by the 1$\sigma$ uncertainty.  Where we have not detected the radio galaxy's millimetre counterpart, we instead mark its 3$\sigma$ upper limit with a red arrow.  Black diamonds show flux density measurements of the synchrotron emission, integrated over the entire radio source.  The solid line is our power-law extrapolation of this emission from the high-frequency radio regime to 1.1 mm.  Where a galaxy's radio core has been detected at more than one radio frequency, its flux density measurements are represented by black squares.  Our power law extrapolation of the radio core SED to 1.1 mm is shown by the dashed black line.  References to the radio measurements from the literature are given in Table 2.}
\end{figure*}

\begin{table*}
\centering
\caption{Extrapolation of a powerlaw from radio to millimetre wavelengths.  The columns are organised as follows.  (1) Name of the radio galaxy; if the radio core has been detected at more than one radio wavelength, it is also included in this table.  (2) Redshift z of the radio galaxy.  (3) Observed flux density of the radio galaxy's 1.1 mm counterpart.  (4) The extrapolated 1.1 mm flux density of the integrated radio source, or radio core where appropriate.  (5) Radio spectral index of the integrated radio source, or the radio core where appropriate.  (6) and (7) The wavelengths of the two radio photometric data points used in the extrapolation.  (8) References to photometric data from the literature which we have used in this paper; references have been abbreviated to the first two letters of the first author's surname and the last two digits of the publication year; the abbreviations are appended to their corresponding full references at the end of this paper.} 
\begin{tabular}{llllllll}
\hline
Source name   & z    & $S_{1.1}$ & $S_{ext}$ & $\alpha_{rad}$ & $\lambda_1$ & $\lambda_2$ & Radio data references\\
              &      & (mJy)     & (mJy)    &                & (cm)        & (cm)      &   \\
(1) & (2) & (3) & (4) & (5) & (6) & (7) & (8)\\
\hline
MRC 2201-555  & 0.51 & 6.1$\pm$0.8 & 0.8 & -1.60 & 3.6 & 6.0 & La81 Wr90 Wr94 Gr94 Bu06 \\
MRC 2008-068  & 0.55 & 8.6$\pm$0.9 & 6.9 & -1.51 & 1.4 & 1.6 & Bo75 Ku81 Wr90 Gr95 Do96 Co98 St98 St05 Ri06a,b He07 \\
MRC 2322-052  & 1.19 & $<$2.1      & 0.2 & -1.65 & 3.6 & 6.0 & Go67 La81 Wr90 Wh92 Gr95 Sl95 Do96 Co98 Co07\\
TXS 2322-040  & 1.51 & 2.3$\pm$0.6 & 0.9 & -1.62 & 3.6 & 6.0 & Wr90 Wh92 Gr95 Do96 Co98 Xi05 Xi06 Li07 \\
MRC 2048-272  & 2.06 & 2.3$\pm$0.8 & 1.1 & -1.15 & 6.2 & 73.4 & La81 Gr94 Do96 Pe00 Se07\\
MRC 0355-037  & 2.15 & 5.0$\pm$0.8 & 0.3 & -1.25 & 21.4 & 73.4 & La81 Wh92 Do96 Co98 Co07 \\
PKS 1138-262  & 2.16 & $<$3.6      & 0.09& -1.84 & 3.7 & 6.4 & La81 Wr90 Gr94 Sl95 Do96 Ca97 Co07 Co98 Re04 \\
PKS 1138-262 core &  &             & 0.02& -1.3 & 3.7 & 6.4 & Ca97 \\
4C +23.56     & 2.48 & $<$1.5      & 0.2 & -1.4 & 3.7 & 6.4 & Pi65 Fa74 Wr90 Gr90 Gr91 Be91 Wh92 Sl95 Do96 Ca97 \\
              &      &             &     &      &     &     & Co07\\
4C+23.56 core &      &             & 0.2 & -0.9 & 3.7 & 6.4 & Ch96 Ca97 \\ 
MRC 2104-242  & 2.49 & 3.7$\pm$0.9 & 0.05& -1.71 & 3.7 & 6.7 & La81 Gr94 Do96 Pe00 Co07\\
MRC 2104-242 core &  &             & 0.0007&-1.6 & 3.7 & 6.7 & Pe00 \\
PKS 0529-549  & 2.58 & 6.4$\pm$0.7 & 1.3 & -1.15 & 1.6 & 3.5 & La81Wr90 Gr94 Wr94 Ma03 Br07 \\
MRC 0316-257  & 3.13 & $<$2.1      & 4.0 & -0.85 & 6.0 & 11.1 & La81 Wr90 Gr94 Do96 Re04 Co07\\
TN J2009-3040 & 3.16 & 3.3$\pm$0.9 & 1.1 & -1.34 & 6.2 & 21.4 & Do96 DeBr00 \\
4C +41.17     & 3.79 & 3.5$\pm$1.0 & 0.05 & -1.65 & 2.0 & 6.1 & Go67 Vi75 Fi85 Ch90 Gr91 Be91 Wh92 Ha93 Do96 Ch96 \\
              &      &             &      &       &     &     & Be99 Gr07 Co07\\
TN J2007-1316 & 3.83 & 2.8$\pm$0.9 & 1.7 & -1.42 & 21.4 & 82.0 & La81 Do96 DeBr00 Re04\\
TN J1338-1942 & 4.10 & $<$5.0      & 0.02 & -1.68 & 3.7 & 6.4 & Do96 Re04 DeBr00 DeBr04 Do07\\
TN J1338-1942 core & &             & 0.004& -1.0 & 3.7 & 6.4 & DeBr00 Pe00\\
TN J0924-2201 & 5.19 & 3.6$\pm$0.9 & 0.01& -1.72 & 6.2 & 21.4 & Do96 vB99 DeBr00 Re04 Ca07\\
\hline
\end{tabular}
\end{table*}

\subsection{Implied star formation rates}
In order to convert our measured flux densities into far-IR luminosities ($L_{FIR}$) and star-formation rates (SFR), we adopt a grey-body emission template with an emissivity index $\beta$ and a dust temperature $T_{dust}$.  This gives

\begin{equation}
M_{dust}=\frac{S_{obs}D_L^2}{(1+z)\kappa_{\nu_{rest}}B_{\nu_{rest}}(T_{dust})}
\end{equation}

\begin{equation}
L_{FIR}=\frac{8 \pi h \kappa_{\nu_{rest}}}{c^2 \nu^{\beta}} \left(\frac{kT_{dust}}{h}\right)^{\beta +4} \Gamma(\beta +4) \zeta(\beta +4) M_{dust}        
\end{equation}

\begin{equation}
SFR=\frac{L_{FIR}}{10^{10} L_{\odot}} M_{\odot} yr^{-1}
\end{equation}

\noindent where $M_{dust}$ is the mass of dust, $S_{obs}$ is the observed flux density, $D_L$ is the luminosity distance, $\kappa_{\nu_{rest}}$ is the mass absorption coefficient of the dust at rest-frame frequency $\nu_{rest}$, and $B_{\nu_{rest}}(T_{dust})$ is the Planck function at $\nu_{rest}$ (e.g. Archibald et al. 2001).  $\Gamma$ is the Gamma function, and $\zeta$ is the Riemann zeta function.  A Salpeter initial mass function has been assumed.  Throughout this paper, we assume $T_{dust}$=40 K, $\beta$=1.5 (after Archibald et al. 2001).  We adopt $\kappa_{375GHz}$=0.15 m$^2$ kg$^{-1}$ (Hildebrand et al. 1983), and extrapolate to other frequencies assuming $\kappa_{\nu}\propto \nu^{\beta}$ (Chini et al. 1986).  

The values we have measured for $S_{1.1}$ imply star formation rates ranging from $<$200 to $\sim$1210 $M_{\odot}$ yr$^{-1}$.  The sample averaged $S_{1.1}$ we obtained from our stacking analysis implies an average SFR$\sim$600 $M_{\odot}$ yr$^{-1}$.  

\section{Comparison with radio and optical properties}

\subsection{Anticorrelation between $S_{1.1}$ and radio source size}
It is interesting to examine what, if any, relationship there might be between mm/sub-mm flux density measurements and the radio source size.  This is motivated by a number of relatively recent observational results suggesting that the presence of intense star-formation in high-z radio-loud galaxies is anticorrelated with the size of the radio source (Best, Longair \& R\"ottgering 1996; Willott et al. 2002; Humphrey et al. 2006).  

For the purpose of this analysis, we define radio source size as the largest projected extent of the radio emission: we have measured the largest angular extent from published radio images, using high spatial resolution images wherever possible.  Whenever possible we use high spatial resolution images (FWHM$\la$0.5\arcsec).  For sources which are not well resolved, or which do not have a radio image published in the literature, we adopt the largest angular size, or the upper limit thereto, listed in the literature.  Angular sizes were converted to physical units of kpc, using the conversion dictated by our preferred cosmology.  

For high-z radio galaxies, the published radio images tend to be sensitive enough only to detect the relatively compact and high surface brightness parts of the radio source, i.e., hotspots and the core.  Therefore, in classical double/triple radio galaxies, the largest angular size typically represents the angular distance between the most distant hotspot, relative to the nucleus, in each of the two jets.  It is important to be aware that it is quite possible that radio emitting structures too faint to be detected lie yet farther from the nucleus (see e.g. Coma A: van Breugel et al. 1985; 3C171: Blundell 1996; B3 J2330+3927 at z=3.087: P\'{e}rez-Torres \& De Breuck 2005).  In this case, our measured radio source sizes would underestimate the true size.  While we clearly cannot ascertain whether the hotspot separations are genuinely representative of the total radio source sizes in our high-z sample, we find it encouraging that in low-z powerful radio galaxies the hotspot separation is usually representative of the total angular extent (see e.g. Black et al. 1992).  

In Fig. 5 we plot $S_{1.1}$ versus the size of the radio emission for our sample.  We find an L-shaped anticorrelation between $S_{1.1}$ and radio size.  The three galaxies with the highest values of $S_{1.1}$ ($\ge$5 mJy: MRC 2201-555; MRC 2008-068; PKS 0529-549) have relatively small radio sources ($\le$22 kpc), while those with relatively large radio sources ($>$22 kpc) all have relatively low $S_{1.1}$ ($\le$5 mJy).  To test the significance of this apparent anticorrelation in our AzTEC sample, we use Cox's proportional hazard model (Isobe, Feigelson \& Nelson 1986), which is able to treat censored data, i.e., upper limits.  Using the null hypothesis that no correlation is present between $S_{1.1}$ and radio size, we obtain a global $\chi ^2$=8.3, which corresponds to a significance level for independence of 0.003.  Thus, the anticorrelation is highly significant.  

An alternative test can be made via Poisson probability theory.  In this case we adopt a null hypothesis such that the population of radio galaxies with relatively large radio sources ($>$50 kpc) has the same fraction of galaxies with $S_{1.1} \ge$5 mJy as does the population of radio galaxies with relatively small radio sources ($\le$50 kpc).  In our sample, there are 3 radio galaxies with $S_{1.1}\ge$5 mJy and D$_{rad} \le$50 kpc; there are 3 with $S_{1.1} <$5 mJy and D$_{rad} \le$50 kpc; there are 10 with $S_{1.1} <$5 mJy and D$_{rad} >$50 kpc; and there are none with $S_{1.1} \ge$5 mJy and D$_{rad} >$50 kpc.  Under the null hypothesis, the probability of counting 0 radio galaxies with $S_{1.1} \ge$5 mJy and D$_{rad} >$50 kpc is 4.5$\times$10$^{-5}$.  Therefore, we reject the null hypothesis, and conclude that the AzTEC radio galaxy sample shows a highly significant anticorrelation between the 1.1 mm flux density and the apparent size of the radio source.  

Shown in Fig. 6 is $S_{1.1}$ versus radio size for the AzTEC counterparts together with HzRGs that have been observed at 850$\mu$m using SCUBA (Archibald et al. 2001; Reuland et al. 2003b, 2004; Stevens et al. 2003).  SCUBA 850$\mu$m flux densities have been scaled by 1/2.5 to extrapolate their 1.1 mm flux densities.  This scaling factor is calculated assuming $T_{dust}$=40 K and $\beta$=1.5 ($\S$3.5).  Where an HzRG has been observed more than once at 1.1 mm and/or 850$\mu$m, we use the most sensitive of the flux density measurements.  Using Cox's proportional hazard model we find this combined dataset does not show a significant anticorrelation between $S_{1.1}$ or SFR and radio size (P=0.18), in agreement with the statistical analysis of Reuland et al. (2004).  However, using Poisson probability theory we find that the absence of HzRGs at $S_{1.1} \ge$2.5 mJy and D$_{rad} >$200 kpc is significant at the 98 per cent level (P=0.02).  

It seems plausible that this reduction in the significance of the trend is due to the different observational methods employed in our AzTEC/ASTE survey of HzRGs, compared to those predominantly used for the several SCUBA/JCMT surveys.  Whereas our maps allows us to reject sources that are nearby to the radio galaxy's position but unlikely to be the counterpart, the SCUBA photometry employed by Archibald et al. (2001) and Reuland et al. (2004) simply obtains the flux density from a region of sky sampled by the telescope beam, and emission from a bright nearby companion MMG may contaminate such measurements, thereby adding scatter to the trend.  

Finally, we point out that the trend discussed above would not change significantly if we were to include only millimetre/sub-millimetre measurements obtained from maps (e.g. from Ivison et al. 2000; Stevens et al. 2003; De Breuck et al. 2004; Greve et al. 2007; and this paper), i.e., excluding measurements obtained from simple SCUBA photometry.  We would still find that, of the radio galaxies with radio sources smaller than 200 kpc, 50 per cent have $S_{1.1}<$5 mJy, while the radio galaxies with radio sources larger than 200 kpc all have $S_{1.1}<$5 mJy.  In $\S$6 we discuss the origin and implications of this result.  

\begin{figure}
\includegraphics{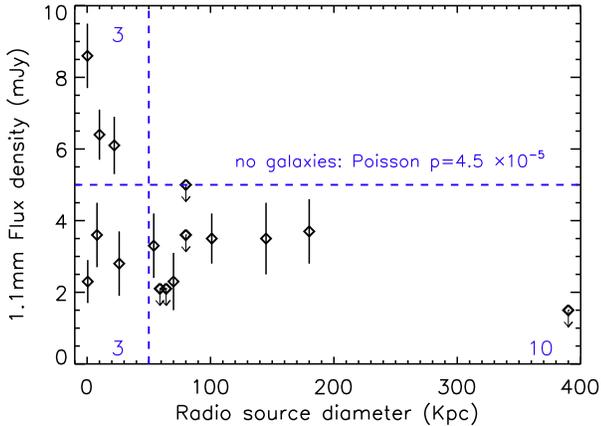}
\vspace{2.35in}
\caption{The anticorrelation between 1.1 mm flux density and radio source diameter in our AzTEC sample.  The dashed lines show the divisions used to calculate the Poisson probability}
\end{figure}

\begin{figure}
\includegraphics{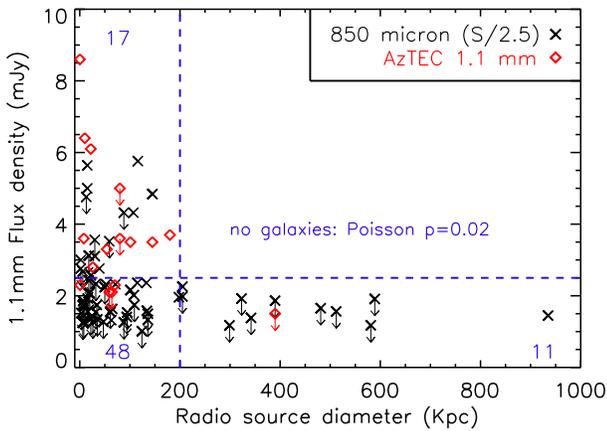}
\vspace{2.35in}
\caption{Similar to Fig. 5, but including also HzRGs with measurements or limits at 850$\mu$m. We have extrapolated 850$\mu$m flux densities to 1.1 mm by multiplying by 1/2.5.  To avoid complicating the figure, error bars are not shown.}
\end{figure}

\begin{figure}
\includegraphics{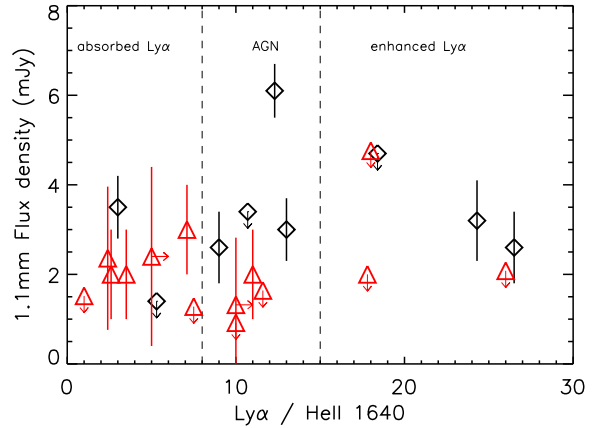}
\vspace{2.35in}
\caption{Flux density at 1.1 mm versus Ly$\alpha$/HeII.  Dashed vertical lines separate three different areas on the plot: Ly$\alpha$/HeII lower than predicted by AGN photoionization models (absorbed); Ly$\alpha$/HeII consistent with AGN photoionization models; and Ly$\alpha$/HeII higher than predicted by AGN photoionization models (enhanced).  Black diamonds show the high-z radio galaxies from our sample which have Ly$\alpha$/HeII measurements in the literature.  Red triangles show high-z radio galaxies from the literature which have both SCUBA 850$\mu$m flux density measurements, which we have scaled by a factor of 0.4 to account for the shorter wavelength, and Ly$\alpha$/HeII measurements, the latter from Villar-Mart{\'{\i}}n et al. (2007).}
\end{figure}

\subsection{The Ly$\alpha$ emission nebulae}
In addition to heating the surrounding dust young stars also produce HII regions, which emit strong UV-optical emission lines.  The strongest of these lines is Ly$\alpha$ $\lambda$1216, and its luminosity is primarily a function of the number of ionizing photons absorbed, and the quantity and geometry of dust.  It is relatively insensitive to the metallicity or the gas density of the HII regions.  Thus, Ly$\alpha$ carries useful information about dust associated with star forming regions.  

At z$\ga$2, Ly$\alpha$ is redshifted into the optical regime, making this line accessible to ground-based telescopes.  Indeed, all 11 of the z$\ge$2.15 radio galaxies in our sample have measurements in the literature for the narrow (FWHM$\la$3000 km s$^{-1}$) Ly$\alpha$ emission.  In table 3, we list the Ly$\alpha$ luminosity for these 11 radio galaxies, measured from narrow band images (e.g. Knopp \& Chambers 1997; Reuland et al. 2003a; Venemans et al. 2007), and/or from spectra with a long-slit placed along the major axis of the radio emission (e.g. De Breuck et al. 2000a).

It is interesting to compare these measured Ly$\alpha$ luminosities against the expected (intrinsic) Ly$\alpha$ luminosity of a massive starburst.  First, we derive an approximate relationship between the Ly$\alpha$ luminosity and SFR or $L_{FIR}$, as
\begin{equation}
L_{Ly\alpha} \sim 10^{44} ({SFR}/{100 M_{\odot} yr^{-1}}) ~erg s^{-1}
\end{equation}

\begin{equation}
L_{Ly\alpha} \sim 10^{44} ({L_{FIR}}/{10^{12} L_{\odot}}) ~erg s^{-1}
\end{equation}

\noindent which simplifies to become

\begin{equation}
L_{Ly\alpha}/{L_{FIR}} \sim 0.03
\end{equation}

\noindent under the assumptions that (i) the starburst is continuous and has an age in the range $\sim$1-5 Myr; (ii) all ionizing photons produced by the stars are absorbed in the surrounding HII region; and (iii) 67 per cent of photo-ionization events lead to the emission of a Ly$\alpha$ photon (Binette et al. 1993).  Of the 8 galaxies in our sample which have been detected in both Ly$\alpha$ emission and in 1.1 mm continuum emission, 3 have values of $L_{Ly\alpha}$/$L_{FIR}$ which are in order-of-magnitude agreement with the expected value $\sim$0.03, and 5 have $L_{Ly\alpha}$/$L_{FIR}$ values more than an order of magnitude lower (Table 3).  

Thus far, we have ignored another luminous source of ionizing radiation which is likely to contribute significantly to the production of Ly$\alpha$ photons -- the active nucleus.  The ratio between Ly$\alpha$ and HeII $\lambda$1640 is sensitive to the SED of the ionizing continuum and, therefore, it can be used to assess the extent to which stellar-photoionized HII regions contribute to the UV-optical emission line spectrum.  Models for pure AGN photoionization predict Ly$\alpha$/HeII = 8 -- 15 (e.g. Humphrey et al. 2008), depending on the spectral slope of the ionizing continuum between 13.6 eV and 54.4 eV.  On the other hand, photoionization by young stars does not result in significant HeII emission relative to Ly$\alpha$.  Thus, a Ly$\alpha$/HeII ratio significantly in excess of 15 would suggest a significant contribution from stellar-photoionized HII regions.  A number of HzRGs, including 3 of our sample (TN J2009-3040, 4C+41.17 and TN J1338-1942), do indeed show such an excess of Ly$\alpha$ emission.  Villar-Mart{\'{\i}}n et al. (2007) examined the Ly$\alpha$/HeII and Ly$\alpha$/CIV ratios of 61 radio galaxies at 1.79$\le$z$\le$4.41, and identified a Ly$\alpha$ excess in 11 of their sample; they calculated that a SFR of $\sim$200 $M_{\odot}$ yr$^{-1}$ would be required to explain the most extreme objects.  

Naively, one might expect there to be a correlation between detection of a Ly$\alpha$ excess and detection of millimetre continuum, since both are thought to be produced by powerful starbursts.  In Fig. 7 we plot $S_{1.1}$ against the Ly$\alpha$/HeII ratio for radio galaxies from our sample which have Ly$\alpha$ and HeII detections.  We also plot radio galaxies which have been observed at 850$\mu$m (from Table 1 of Villar-Mart{\'{\i}}n et al. 2007), under the assumption that $S_{1.1}$=$S_{850}$/2.5.  Contrary to initial expectations, we find no trend between $S_{1.1}$ and Ly$\alpha$/HeII.   The possible origin of this result is discussed in $\S$6.

\begin{table*}
\centering
\caption{Radio and optical properties of the radio galaxies.  A dash (--) indicates that a quantity has not yet been measured, or is not available in the literature.  Columns are as follows.  (1) Source name.  (2) Source redshift.  (3) The 1.1 mm flux density of the millimetre counterpart, measured from our AzTEC data.  (4) The spectroscopic Ly$\alpha$ luminosity in units of $10^{44}$ erg s$^{-1}$.  (5) Ly$\alpha$ luminosity in units of $10^{44}$ erg s$^{-1}$ measured from narrow-band images.  (6) The flux ratio Ly${\alpha}$/HeII (7) Ratio of the Ly$\alpha$ luminosity to the far-infrared luminosity; the theoretically expected value is $\sim$0.03; Ly$\alpha$ luminosities measured from narrow band images are used where available, otherwise luminosities measured from long-slit spectra are used. (8) Projected size of the radio source (D$_{radio}$).  (9) References for the size of the radio source, and for the luminosities of the Ly$\alpha$ and HeII emission lines; references have been abbreviated to the first two letters of the first author's surname and the last two digits of the publication year, and the abbreviations are appended to the full bibliographic references at the end of this paper.} 
\begin{tabular}{lllllllll}
\hline
Source name   & z    & $S_{1.1}$ & $L_{Ly\alpha}$(slit) &  $L_{Ly\alpha}$(image) & Ly${\alpha}$/HeII & $L_{Ly\alpha}$/$L_{FIR}$ & D$_{radio}$ & Refs \\
              &      & (mJy)       & (10$^{44}$ erg s$^{-1}$) & (10$^{44}$ erg s$^{-1}$) &      &          & (kpc)  &  \\
(1)           & (2)  & (3)       & (4)  & (5)  & (6)      & (7)   & (8) & (9) \\
\hline
MRC 2201-555  & 0.51 & 6.1$\pm$0.8 & --   & --  & --   &   --     & 22   & Bu06 \\
MRC 2008-068  & 0.55 & 8.6$\pm$0.9 & --   & --  & --   &   --     & 0.2  & Je00 \\
MRC 2322-052  & 1.19 & $<$2.0      & --   & --  & --   &   --     & 64   & Be99 \\
TXS 2322-040  & 1.51 & 2.3$\pm$0.6 & --   & --  & --   &   --     & 0.5  & Li07 \\
MRC 2048-272  & 2.06 & 2.3$\pm$0.8 & --   & 0.65& --   & 0.0035   & 70   & Pe00 Ve07\\
MRC 0355-037  & 2.15 & 5.0$\pm$0.8 & 0.40 & --  & 3.0  & 0.0010   & 101  & Go05 R\"o97 \\
PKS 1138-262  & 2.16 & $<$3.4      & 1.7  & 25  & 10.7 & $>$0.025 & 80   & Ca97 R\"o97 Ve07 \\
4C +23.56     & 2.48 & $<$1.4      & 0.40 & 2.1 & 5.3  & $>$0.018 & 390  & Ca97 Ch96 R\"o97 Ci98 Kn97 \\
MRC 2104-242  & 2.49 & 3.7$\pm$0.9 & 2.9  & --  & 13   & 0.010    & 180  & Pe00 Vi99 Ov01\\
PKS 0529-549  & 2.58 & 6.4$\pm$0.7 & 0.41 & --  & 12.3 & 0.00084  & 10   & Br07 R\"o97 \\
MRC 0316-257  & 3.13 & $<$1.9      & 0.21 & 0.7 & --   & $>$0.0047& 59   & Ca97 At98 Ve07\\
TN J2009-3040 & 3.16 & 3.3$\pm$0.9 & 1.4  & 3.0 & 24.3 & 0.013    & 54   & Bo07 DeBr00 Ve07 \\
4C +41.17     & 3.79 & 3.5$\pm$1.0 & 2.1  & 13  & 26.5 & 0.054    & 145  & Ch90 Ch96 R\"o97 Re03 \\
TN J2007-1316 & 3.83 & 2.8$\pm$0.9 & 0.26 & --  & 9.0  & 0.0014   & 26   & Bo07 DeBr00 \\
TN J1338-1942 & 4.10 & $<$4.7      & 1.7  & 4.5 & 18.4 & $>$0.014 & 80   & Pe00 DeBr99 DeBr01 Ve07 \\
TN J0924-2201 & 5.19 & 3.6$\pm$0.9 & 0.10 & 0.15& --   & 0.00074& 8    & vB99 DeBr01 Ve07 \\
\hline
\end{tabular}
\end{table*}

\begin{figure}
\includegraphics{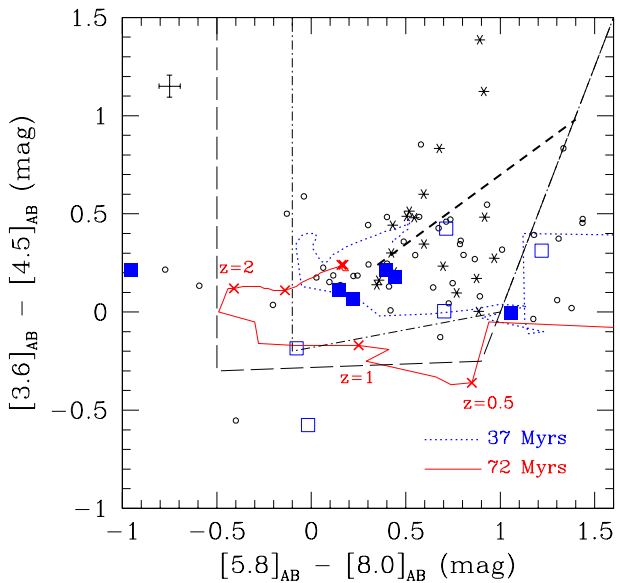}
\vspace{3.14in}
\caption{{\it Spitzer} IRAC [3.6]-[4.5] versus [5.8]-[8.0] colour-colour diagram for the radio galaxies in our sample which have been detected in all 4 bands (adapted from Stern et al. 2005).  Solid squares are HzRGs for which we have detected a probable 1.1 mm counterpart; open squares are HzRGs for which we have not detected a probable 1.1 mm counterpart.  The small open circles are the HzRG sample of Seymour et al. (2007).  Asterisks (*) show the infrared QSOs with power law spectra from Lacy et al. (2004) and Mart\'{i}nez-Sansigre et al. (2008).  The dot-dashed line indicates the AGN region proposed by Stern et al. (2005).  The long-dashed black line shows the MMG counterpart region proposed by Yun et al. (2008).  The dotted line and the solid line show the redshift evolution colour-colour track for a starburst age of 37 and 72 Myr, respectively, based on the theoretical starburst spectral energy distribution models of Efstathiou et al. (2000); crosses (x) along the 72 Myr track mark z = 0.5, 1, 2, 3, 4, and 5.  The thick short-dashed line represents power-law spectrum sources having $\alpha$=0.3-1.0, where $S_v \propto v^{-\alpha}$.}
\end{figure}

\begin{figure}
\includegraphics{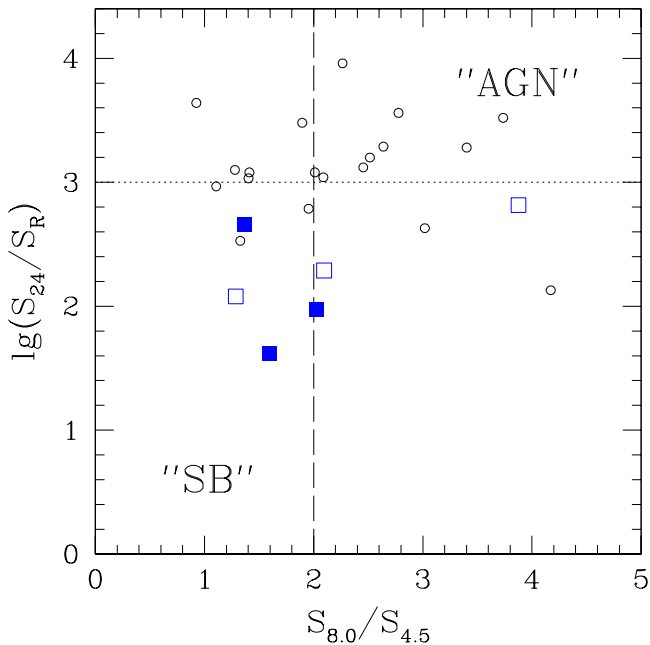}
\vspace{3.54in}
\caption{Log $S_{24}/S_{R}$ versus $S_{8.0}/S_{4.5}$ for the radio galaxies in our sample which have been detected in the {\it Spitzer} MIPS 24 $\mu$m channel, in the IRAC 4.5 and 8.0 $\mu$m channels, and in the optical R-band.  Filled squares show radio galaxies for which we have detected a probable counterpart at 1.1 mm; open squares show radio galaxies for which we have not detected a probably 1.1 mm counterpart.  The small circles are the HzRG sample of Seymour et al. (2007) as in Figure~8.  Note that $S_{R}$ is the flux density corresponding to the R or r magnitude of the radio galaxy.  References to the optical photometry are as follows.  MRC 2201-555: Bu06.  MRC 0355-037: Go05.  PKS 1138-262: Pe97  MRC 0316-257: Mc96.  4C +41.17: Mi92.  TN J1338-1942: DeBr02.  Starburst galaxies (SB) are expected to lie at Log $S_{24}/S_{R}<3$, $S_{8.0}/S_{4.5}<$2, whereas active galaxies are expected to lie above and/or to the right of this region of colour space (see Pope et al., 2008; Dey et al. 2008; Fiore et al. 2008).}
\end{figure}

\begin{figure}
\includegraphics{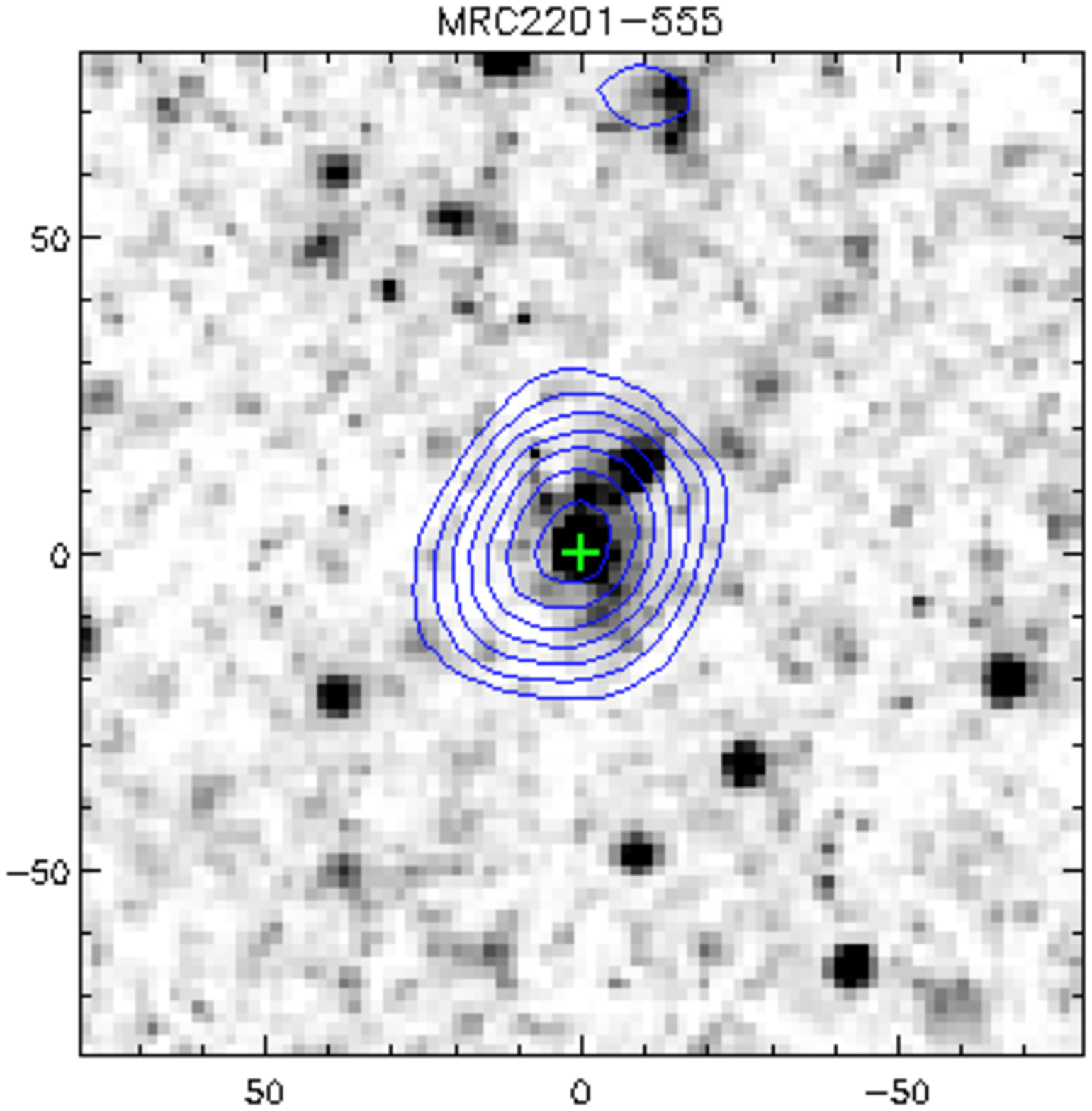}
\includegraphics{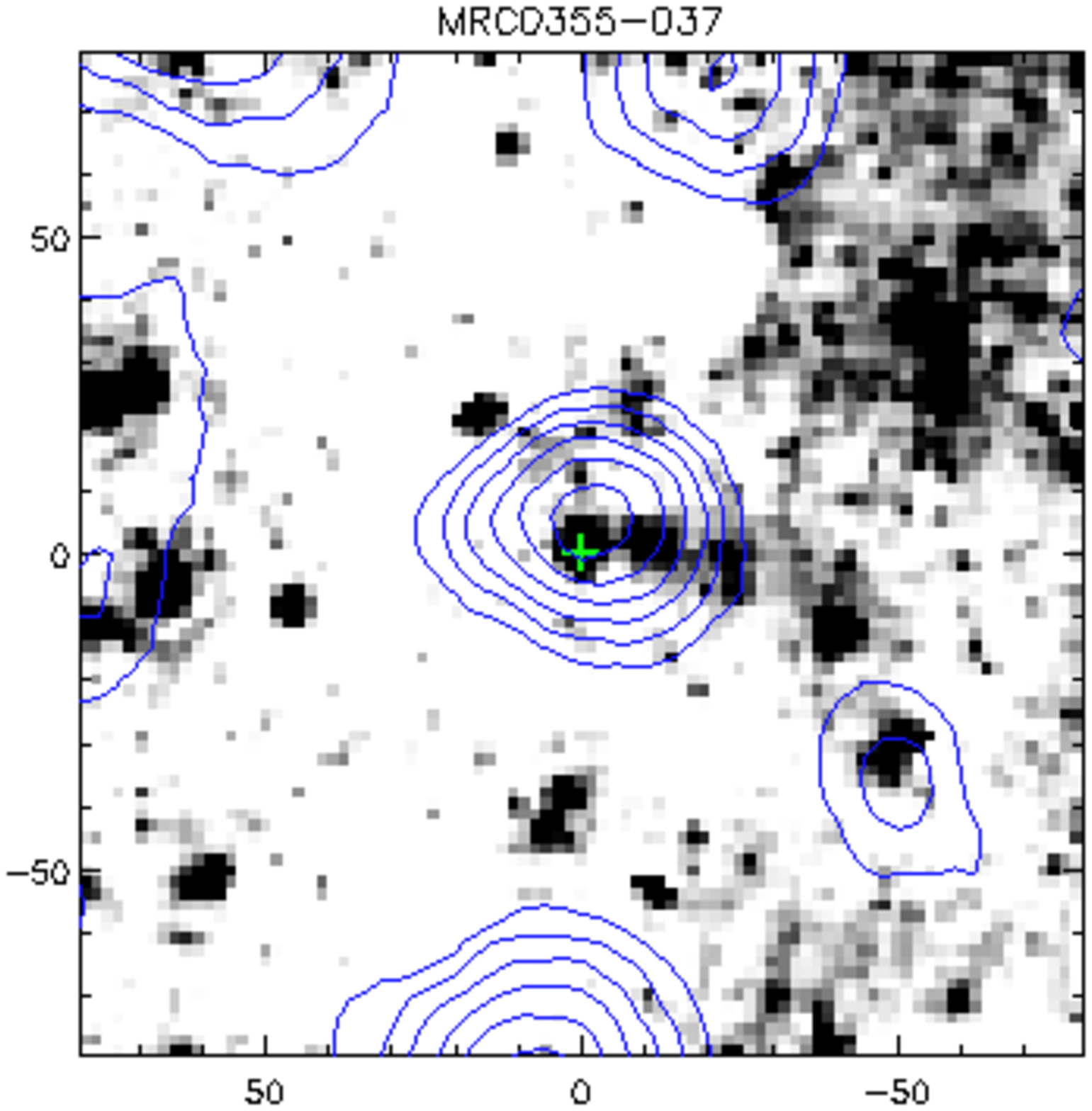}
\includegraphics{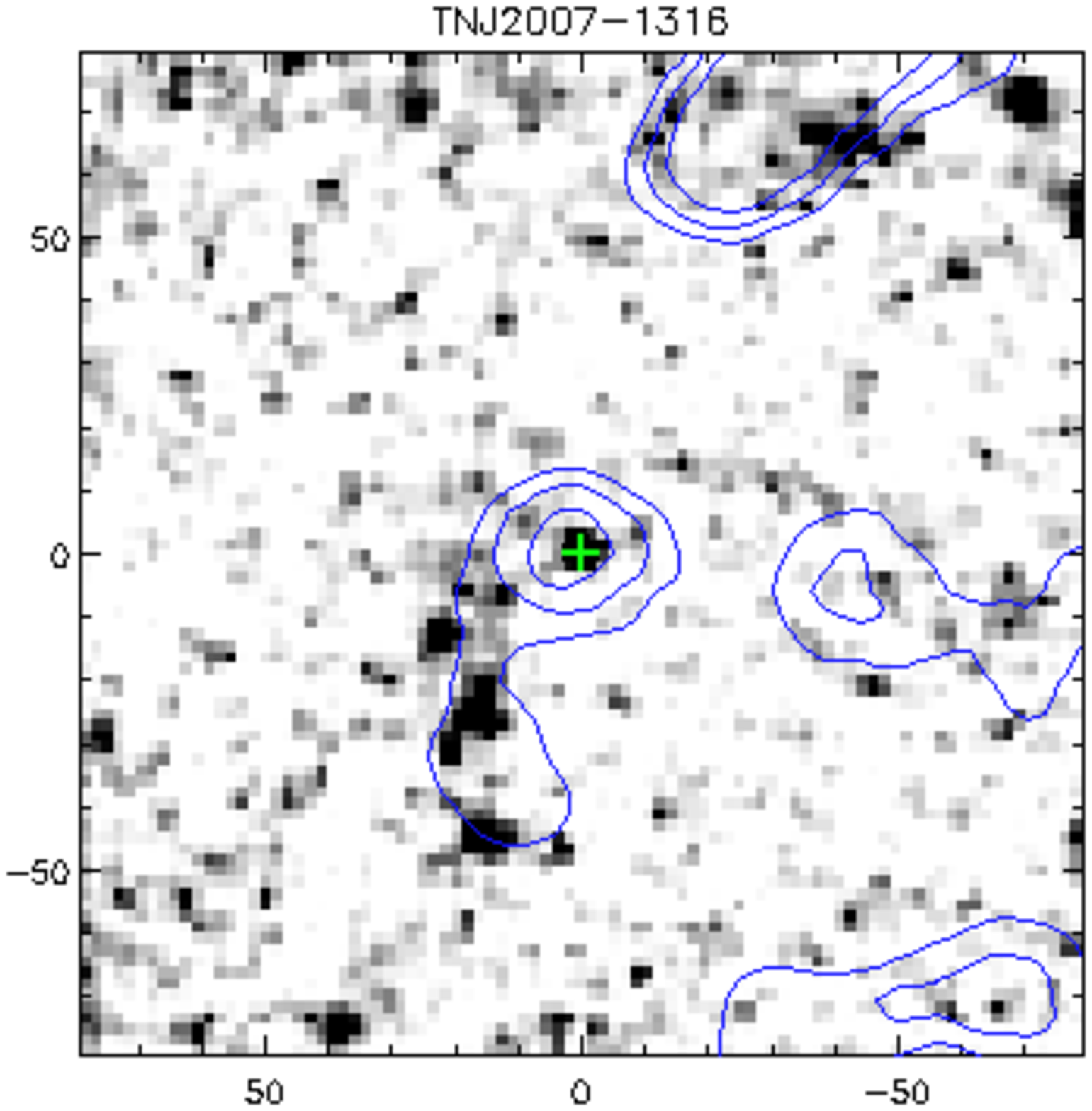}
\vspace{8.48in}
\caption{Postage stamps of the 24$\mu$m emission (greyscale) with contours of the 1.1 mm emission overlayed.  For MRC 2201-555, the AzTEC contours are the same as in Fig. 1.  For MRC 0355-037, the contours start at 1$\sigma$ and increase by 1$\sigma$.  For TN J2007-1316, the contours are 1.7, 2.2, 2.7 $\sigma$.  North and East are at the top and left of the images, respectively.  The cross (+) marks the position of the active galaxy.}
\end{figure}

\section{Analysis of the mid-IR spectral energy distribution}
In Fig.~8 we show the {\it Spitzer} IRAC [3.6]-[4.5] versus [5.8-8.0] colour-diagram, adapted from Stern et al. (2005), for all HzRGs in our sample that are detected in all 4 IRAC channels.  Most of our target HzRGs, as well as those studied by Seymour et al. (2007), are found within the region of red rest-frame optical and near-IR color characterisitic of hot dust, overlapping with the distribution of the {\it Spitzer}-selected power-law QSOs of Lacy et al. (2004) and Mart\'{i}nez-Sansigre et al. (2008).  While the colours of our target HzRGs are marginally consistent with colours expected for active galaxies, those detected by AzTEC (filled squares) appear near the edges of the distribution, along the model tracks of young dust obscured stellar clusters modeled by Efstathiou et al. (2000).  This is also the area where the majority of submillimetre-bright sources are clustered (Yun et al. 2008, 2011), and these AzTEC detected HzRGs display rest-frame near-IR properties more like SMGs than other HzRGs.     

The diagram shown in Fig.~9 aims to distinguish active galaxies from starburst systems using two different diagnostic criteria proposed recently by different groups: (1) $S_{24}/S_R >$ 1000 (Dark Optical Galaxies, or DOGs: Dey et al. 2008; Fiore et al. 2008); and (2) $S_{8.0}/S_{4.5} >$ 2 (Ivison et al. 2004; Pope et al. 2008).  The DOGs were first identified in the study of a population of $z\sim2$ infrared bright galaxies that are extremely faint in the optical bands, and Fiore et al. (2008) concluded based on their X-ray analysis that as many as 80\% of these dust-obscured galaxies host a Compton-thick AGN.  The latter criterion was initially developed to exploit the presence of PAH features and a power-law AGN spectral shape in the rest-frame near-IR, similar to the information used in Fig.~8.
 
In this diagram, the majority of the HzRGs detected in MIPS 24 \micron\ by Seymour et al. (2007) appear above the DOGs line of $S_{24}/S_R >$ 1000 (dotted line), suggesting warm dust and a heavily obscured AGN are present in these HzRGs.  All six of our target HzRGs detected in the MIPS 24 \micron\ bands are below the dotted line and are on average bluer than the others.  In fact, the three AzTEC-detected HzRGs are among the bluest HzRGs.  The AzTEC detected HzRGs also show a flatter $S_{8.0}/S_{4.5}$ flux ratio than the rest of the HzRGs.  Taken together, these rest-frame optical and near- and mid-IR diagnostic colours suggest that the AzTEC-detected HzRGs have bluer colours of star-forming galaxies than those of dust-obscured AGNs with characteristic hot dust emission commonly associated with other HzRGs.  We note that the overall rest frame optical colours of even the most dusty star forming galaxies, such as ultraluminous infrared galaxies in the local universe, are surprisingly blue because young stellar populations distributed throughout the galaxy dominates the overall colour and luminosity (see Chen et al. 2010).

In summary, the HzRGs which have an AzTEC counterpart also have mid-infrared spectral energy distributions that are consistent with dust-obscured starburst systems.

\begin{table*}
\centering
\caption{Photometric measurements from the {\it Spitzer} observations of our sample of high-z active galaxies.} 
\begin{tabular}{lllllll}
\hline
Source name   & z    & S$_{3.6}$ & S$_{4.5}$ & S$_{5.8}$ & S$_{8.0}$ & S$_{24}$ \\
              &      & ($\mu$Jy) & ($\mu$Jy) & ($\mu$Jy) & ($\mu$Jy) & ($\mu$Jy)  \\
\hline
MRC 2201-555  & 0.51 & 377$\pm$6     & 401$\pm$7     & 522$\pm$22   & 640$\pm$27   & 1547$\pm$255 \\
MRC 2322-052  & 1.19 & 147$\pm$2     & 124$\pm$2     & 96$\pm$5     & 90$\pm$5     & --           \\
TXS 2322-040  & 1.51 & --            & --            & --           & --           & 51$\pm$15    \\
MRC 2048-272  & 2.06 & 61$\pm$2      & 71$\pm$3      & 91$\pm$14    & 73$\pm$20    & 268$\pm$61   \\
MRC 0355-037  & 2.15 & 32$\pm$1      & 35$\pm$1      & 42$\pm$4     & 48$\pm$5     & 812$\pm$136  \\
PKS 1138-262  & 2.16 & 353$\pm$5     & 522$\pm$9     & 1048$\pm$44  & 2024$\pm$85  & 5111$\pm$824 \\
4C +23.56     & 2.48 & 60$\pm$1      & 80$\pm$1      & 164$\pm$7    & 505$\pm$21   & 5409$\pm$876 \\
MRC 2104-242  & 2.49 & 27$\pm$6      & 33$\pm$7      & 37$\pm$22    & $<$75        & 876$\pm$255  \\
PKS 0529-549  & 2.58 & 46$\pm$2      & 55$\pm$2      & 68$\pm$12    & 102$\pm$14   & 632$\pm$110  \\
MRC 0316-257  & 3.13 & 27.1$\pm$0.5  & 27.2$\pm$0.6  & 30$\pm$2     & 57$\pm$3     & 661$\pm$118  \\
TN J2009-3040 & 3.16 & --            & --            & --           & --           & 18$\pm$4     \\
4C +41.17     & 3.79 & 21.3$\pm$0.6  & 26.0$\pm$0.7  & 36$\pm$3     & 52$\pm$4     & 467$\pm$82   \\
TN J2007-1316 & 3.83 & 62$\pm$2      & 61$\pm$3      & 65$\pm$14    & 172$\pm$20   & 594$\pm$102  \\
TN J1338-1942 & 4.10 & 21.8$\pm$0.8  & 12.8$\pm$0.6  & 17$\pm$3     & 17$\pm$3     & 283$\pm$61   \\
TN J0924-2201 & 5.19 & 6.7$\pm$0.3   & 10$\pm$3      & 9.7$\pm$0.9  & $<$37        & 217$\pm$68   \\
SDSS J1030+0524& 6.28& 86$\pm$1      & 109$\pm$2     & 92$\pm$5     & 109$\pm$7    & 523$\pm$92   \\
\hline
\end{tabular}
\end{table*}

\subsection{24$\mu$m images}
In Fig.~10 we show 160\arcsec$\times$160\arcsec postage stamps of the 24$\mu$m emission, with contours of the 1.1 mm emission overlayed, in order to illustrate interesting morphological features around 3 of the radio galaxies.  A detailed analysis of clustering in the full 170-300 arcmin$^2$ fields (as opposed to the 7 arcmin$^2$ `postage stamps' considered herein) of our sample of 17 high-z active galaxies will be presented in a future paper (Zeballos et al., in preparation).

\subsubsection{MRC 2201-555}
The 24 $\mu$m MIPS image shows an arc of emission running through the position of the radio galaxy, with a total spatial extent of $\sim$38\arcsec or 230 kpc.  The NW portion of the arc takes the form of a relatively bright source $\sim$16\arcsec (100 kpc) to the NW, which is linked to the radio galaxy by a bridge of lower surface brightness emission.  This structure is too compact to resolve at the resolution of our AzTEC/ASTE observation.  

\subsubsection{MRC 0355-037}
The MIPS image shows an arc comprised of several 24 $\mu$m sources extending $\sim$60\arcsec ($\sim$500 kpc) to the WSW from the position of the radio galaxy.  At or near to the end of this arc, our AzTEC map shows a peak which has S/N=2.6, after correcting for the negative sidelobes of radio galaxy's AzTEC counterpart.  

\subsubsection{TN J2007-1318}
An arc of several 24 $\mu$m sources extends $\sim$50\arcsec ($\sim$360 kpc) SSE from the radio galaxy.  Similarly, in our AztEC image there is arc of 1.1 mm emission, cospatial with the 24 $\mu$m arc, with peak S/N=2.4 after correcting for the negative sidelobes of radio galaxy's 1.1 mm counterpart.  

\section{Discussion}
In the process of this study, we have identified a statistically significant anti-correlation between the size of the radio source and the brightness of the millimetre continuum emission (or SFR).  This is by no means the first investigation into the possible relationships between the size of a radio source and the other properties of radio galaxies; numerous relationships have been found previously between the observed size of powerful radio sources and various UV, optical and infrared properties.  Best, Longair \& R\"ottgering (1996) found a morphological evolution with radio size in their sample of eight 3CR radio galaxies at 1$<$z$<$1.3.  For galaxies with relatively smaller radio sources, their {\it HST} images reveal a string of bright knots tightly aligned with the radio emission.  The galaxies with relatively larger radio sources, while still showing the alignment effect (e.g. Chambers, Miley \& van Breugel 1987; McCarthy et al. 1987), are generally more compact and contain fewer bright optical knots.  Furthermore, galaxies with smaller radio sources, the extended emission line region (EELR) tend to have a relatively smaller observed spatial extent, as measured with Ly$\alpha$ (van Ojik et al. 1997).  The EELR of these galaxies also show more extreme emission line kinematics, have brighter UV-optical emission lines, and ratios between emission lines that imply they have lower ionization states (Best, R\"ottgering \& Longair 2000; Inskip et al. 2002; Humphrey et al. 2006).  In addition, the polarization of the rest-frame UV continuum emission is correlated with radio source size (Sol\'orzano-I\~narrea et al. 2004), suggesting that the ratio between scattered light and starlight is relatively higher in radio galaxies with larger radio sources.  The anticorrelation between 850 $\mu$m luminosity and UV continuum polarization found by Reuland et al. (2004) appears to confirm that the UV continuum polarization is indeed strongly dependent on the rate of star formation.  Furthermore, HzRGs showing anomalously high values of Ly$\alpha$/HeII ($\ga$15), likely due to the presence of a powerful starburst ($\sim$100 $M_{\odot}$ yr$^{-1}$) tend to have smaller radio sources than those with 'normal' or 'low' values (Villar-Mart\'{i}n et al. 2007).  These radio-optical trends are commonly supposed to be the result of (a) interaction between the radio source and the host galaxy, (b) slowing of radio source expansion in radio galaxies that are in relatively denser environments, or (c) age effects of the radio source.  

A previously unanswered question was whether the star formation rate is intrinsically higher in HzRGs with smaller radio sources, or whether instead the UV-optically derived SFR merely appears higher due to some difference in dustiness, geometry, etc.  Observations in the millimetre/sub-mm regime give us a new perspective from which we may address this question, since the observations are sensitive to UV-optically obscured (rather than unobscured) star forming regions.  Assuming that the millimetre/sub-mm continuum is dominated by thermal emission from starburst-heated dust, as we believe to be the case, then our anticorrelation between $S_{1.1}$ and radio size leads us to conclude that HzRGs with relatively smaller radio sources do indeed tend to have intrinsically higher SFR.  

\begin{figure}
\includegraphics{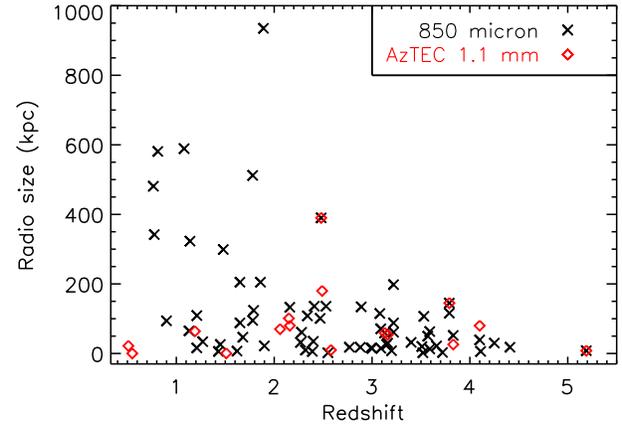}
\vspace{2.35in}
\caption{Radio size as a function of z.  In the interest of clarity, error bars are not shown.  Note the clear anticorrelation between these two parameters, which complicates interpretation of any observed evolution of properties with z (see text). }
\end{figure}

In light of this conclusion, it seems natural to now question whether the increase in the mm/sub-mm detectability of HzRGs at z$\ge$2.5 is due to a genuine redshift evolution in the rate or mode of star formation in the host galaxies, as has commonly been supposed (Archibald et al. 2001; Reuland et al. 2004).  Indeed, we propose an alternative hypothesis: that there is a bias towards identifying HzRGs with smaller radio sources at higher z (Fig. 12) because these sources are associated with more luminous narrow emission line regions (e.g. Best, R\"ottgering \& Longair 2000).  And since smaller radio sources typically have more luminous mm/sub-mm emission (Figs. 5 \& 6), we propose that this translates into a bias in favour of identifying HzRGs with significantly more luminous mm/sub-mm emission at higher z.  Identifying and studying the `missing', line-weak HzRGs will be crucial for testing this hypothesis (see Reuland et al. 2003b; Reuland 2005; Miley \& De Breuck 2008).  

It is not known whether the observed size of the radio source is determined primarily by the density of the interstellar/intergalactic medium through which the source propagates, or whether the size instead determined primarily by the age of the radio source.  If we were to assume the latter, then we would be able to gauge the lifetime of luminous, dusty star formation episodes (S$_{1.1}>$2.5 mJy; SFR$>$500 M$_{\odot}$ yr$^{-1}$) closely associated with HzRGs.  In Fig. 5, all of the luminous millimetre counterparts are associated with HzRGs that have radio source diameters of d$<$200 kpc: we adopt d$\sim$200 kpc as the approximate stage in the growth of the radio source at which the luminous millimetre counterparts fade to S$_{1.1}<$2.5 mJy (SFR$<$500 M$_{\odot}$ yr$^{-1}$).  Assuming a conservative expansion speed of $\sim$0.01c for the radio source (e.g. Best et al. 1995), this diameter implies an age of $\sim$30 Myr.  A similar age was estimated by Willott et al. (2002) for a smaller sample of high-z, radio-loud quasars.  However, we must be cautious about generalising this to obscured star formation in non-active galaxies (i.e., to sub-mm galaxies in general): HzRGs contain two powerful sources of feedback, the radio source and the accretion disc, both of which are able to quench star formation and, perhaps, to ignite it.  Concievably, feedback of this kind could either lengthen or shorten the lifetimes of star formation activity in HzRGs, and could mean that lifetimes of obscured star formation activity determined for HzRGs are not necessarily representative of those for millimetre/sub-millimetre galaxies in general.  

We have also investigated whether there might be a correlation between the millimetre flux density and the Ly$\alpha$/HeII ratio, which is expected to be sensitive to SFR also.  While Ly$\alpha$ can originate from stellar photoionized HII regions and from AGN-ionized nebulae, the HeII line is expected to come predominantly from nebulae that are ionized by the AGN.  Though both Ly$\alpha$/HeII and the millimetre flux density are expected to be correlated with SFR, and therefore with each other, no significant trend was found between them.  Furthermore, the luminosity of the Ly$\alpha$ emission is often more than an order of magnitude lower than one would have expected given the SFR derived from the 1.1 millimetre flux density (we estimate $L_{Ly\alpha}$/$L_{FIR}\sim$0.0007-0.054, while the expected value is $\sim$0.03).  This may mean that one (or both) of the millimetre and Ly$\alpha$ emission is an inaccurate tracer of the SFR in HzRGs.  This could be due to destruction of Ly$\alpha$ photons by the dust associated with this kind of star formation activity.

Thus far, we have tacitly assumed that the sources we identify as 1.1 mm counterparts are the radio galaxy hosts themselves.  However, it is important to be aware that the spatial resolution of our observations, together with the signal to noise ratios of the detections, has resulted in 1$\sigma$ positional uncertainties of $\sim$2\arcsec-16\arcsec for our 1.1 mm counterparts.  This means that the sources we identify as 1.1 mm counterparts could in actual fact be offset by tens or even hundreds of kpc from the galactic nucleus, which would place them in the outskirts or outside of the host galaxy.  This may be the case for TN J0924-2201, which is undetected in SCUBA photometry (Reuland et al. 2004), but which according to our criteria has a 1.1 mm counterpart in our AzTEC map ($\S$3.2) -- this suggests that the AzTEC source is not the actual counterpart, but is instead a MMG several tens of kpc from the radio galaxy.  Also of relevance in this context are results from a FWHM=2\arcsec spatial resolution SMA study of the z$=$3.79 radio galaxy 4C +60.07 (Ivison et al. 2008): the luminous sub-mm emission, previously assumed to be centred on the active galaxy's host, was resolved into two discrete sub-mm sources with offsets of 1\arcsec ($\sim$10 kpc) and 4\arcsec ($\sim$30 kpc) from the position of the radio core.  These two results provide evidence, although anecdotal in nature, that the millimetre/sub-mm emission associated with high-z radio galaxies may not always be spatially coincident with the host galaxy.  Unfortunately, with currently available facilities high resolution observations of the kind presented by Ivison et al. (2008) are costly and are sensitive only the very brightest of MMGs.  The next generation of millimetre/sub-mm facilities, such as the 50 m Large Millimetre Telescope or the Atacama Large Millimetre Array, will be needed to resolve this issue.  

Nevertheless, we can state that all 17 of the active galaxies in our sample are associated with one or more MMG, and thus we conclude that high-z (radio-loud) active galaxies are beacons for finding MMGs.

\section{Summary}
We have presented 1.1 mm imaging observations of the 7 arcmin$^2$ fields around 16 powerful radio galaxies at 0.5$<$z$<$5.2 and a radio quiet quasar at z=6.3.  With these new observations, we have more than doubled the number of high-z radio galaxies which have been imaged at millimetre/sub-millimetre wavelengths.  For all 17 active galaxies, we detect at least one associated millimetric source, which shows that high-z (radio-loud) active galaxies are beacons for finding millimetre/sub-millimetre galaxies at high-z.  We identify likely millimetric counterparts for 11 radio galaxies, and these have 1.1 mm flux densities ranging between $<$1.4 mJy (3$\sigma$) and 8.6$\pm0.9$ mJy, with implied star formation rates of $<$200 to $\sim$1300 $M_{\odot}$ yr$^{-1}$.  After stacking our images at the position of the radio galaxies, including those for which we did not detect the millimetric counterpart, we derive an average flux density of 3.4$\pm$0.2 mJy.  

For the high-z radio galaxies, we have identified an anticorrelation between 1.1 mm flux density (or star formation rate) and the projected linear size of the radio source.  From this result, we have concluded that smaller radio sources are associated with more powerful (obscured) starbursts.  

We have also presented images and photometry from new and archival {\it Spitzer} IRAC and MIPS observations.  We find that the mid-infrared spectral energy distributions of the radio galaxies tend to be bluer than those of {\it Spitzer}-selected high-z active galaxies (Lacy et al. 2004; Mart\'{i}nez-Sansigre et al. 2008).  In particular, the radio galaxies which we have detected at 1.1 mm have mid-infrared colours consistent with obscured star formation.  We have also identified three $\ge$100 kpc arcs of 24$\mu$m sources of which three radio galaxies appear to be part; for the two arcs at z$>$2, we note a spatial correlation between the 24 $\mu$m and the 1.1 mm emission.  

\section*{Acknowledgments}
We thank the referee, Carlos De Brueck, for suggestions which improved this paper, for making published MAMBO maps of 4C+41.17 and TN J1338-1942 available to us, and for providing us with an unpublished radio image of TN J2007-1316 which we used to determine its radio source position angle (28$^{\circ}$) and the position of its radio core (see Table 1).  We also thank Tom Downes for his important contribution to the AzTEC data pipeline software.  AH acknowledges useful discussions with Alfredo Monta\~na and Emmaly Aguilar, and financial support from both CONACyT and INAOE.  This research was partially supported by CONACyT project 39953F.  This work has also been supported in-part by NSF Grants 0907952 and 0838222.  The James Clerk Maxwell Telescope is operated by The Joint Astronomy Centre on behalf of the Science and Technology Facilities Council of the United Kingdom, the Netherlands Organisation for Scientific Research, and the National Research Council of Canada.  The ASTE project is driven by the Nobeyama Radio Observatory (NRO), a branch of the National Astronomical Observatory of Japan (NAOJ), in collaboration with the University of Chile and Japanese institutions including the University of Tokyo, Nagoya University, Osaka Prefecture University, Ibaraki University, and Hokkaido University.

\section*{References}

Archibald E.~N., Dunlop J.~S., Hughes D.~H., Rawlings S., Eales S.~A., Ivison R.~J., 2001, MNRAS, 323, 417 
\\
Athreya R.~M., Kapahi V.~K., McCarthy P.~J., van Breugel W., 1998, A\&A, 329, 809 (At98)
\\
Austermann J.~E., et al., 2010, MNRAS, 401, 160 
\\
Becker R.~H., White R.~L., Edwards A.~L., 1991, ApJS, 75, 1 (Be91)
\\
Benford D.~J., 1999, Thesis, California Institute of Technology (Be99)
\\
Bertin E. \& Arnouts S., 1996, A\&AS, 117, 393
\\
Best P.~N., Bailer D.~M., Longair M.~S., Riley J.~M., 1995, MNRAS, 275, 1171
\\
Best P.~N., Longair M.~S., Rottgering H.~J.~A., 1996, MNRAS, 280, L9
\\
Best P.~N., R{\"o}ttgering H.~J.~A., Lehnert M.~D., 1999, MNRAS, 310, 223 (Be99)
\\
Best P.~N., R{\"o}ttgering H.~J.~A., Longair M.~S., 2000, MNRAS, 311, 23 
\\
Binette L., Wang J., Villar-Martin M., Martin P.~G., Magris C.~G., 1993, ApJ, 414, 535 
\\
Black A.~R.~S., Baum S.~A., Leahy J.~P., Perley R.~A., Riley J.~M., Scheuer P.~A.~G., 1992, MNRAS, 256, 186 
\\
Blundell K.~M., 1996, MNRAS, 283, 538
\\
Bolton J.~G., Shimmins A.~J., Wall J.~V., 1975, AuJPA, 34, 1 (Bo75)
\\
Bornancini C.~G., De Breuck C., de Vries W., Croft S., van Breugel W., R{\"o}ttgering H., Minniti D., 2007, MNRAS, 378, 551 (Bo07)
\\
Broderick J.~W., De Breuck C., Hunstead R.~W., Seymour N., 2007, MNRAS, 375, 1059 (Br07)
\\
Burgess A.~M., Hunstead R.~W., 2006, AJ, 131, 114 (Bu06)
\\
Carilli C.~L., Roettgering H.~J.~A., van Ojik R., Miley G.~K., van Breugel W.~J.~M., 1997, ApJS, 109, 1 (Ca97)
\\
Carilli C.~L., Wang R., van Hoven M.~B., Dwarakanath K., Chengalur J.~N., Wyithe S., 2007, AJ, 133, 2841 (Ca07)
\\
Cimatti A., di Serego Alighieri S., Vernet J., Cohen M., Fosbury R.~A.~E., 1998, ApJ, 499, L21 (Ci98)
\\
Chambers K.~C., Miley G.~K., van Breugel W., 1987, Natur, 329, 604 
\\
Chambers K.~C., Miley G.~K., van Breugel W.~J.~M., 1990, ApJ, 363, 21 (Ch90)
\\
Chambers K.~C., Miley G.~K., van Breugel W.~J.~M., Bremer M.~A.~R., Huang J.-S., Trentham N.~A., 1996, ApJS, 106, 247 (Ch96)
\\
Chen, Y., Lowenthal, J.~D., Yun, M.~S., 2010, ApJ, 712, 1385
\\
Chini R., Kr\"ugel E., Kreysa E., 1986, A\&A, 167, 315
\\
Cohen A.~S., Lane W.~M., Cotton W.~D., Kassim N.~E., Lazio T.~J.~W., Perley R.~A., Condon J.~J., Erickson W.~C., 2007, AJ, 134, 1245 (Co07)
\\
Condon J.~J., Cotton W.~D., Greisen E.~W., Yin Q.~F., Perley R.~A., Taylor G.~B., Broderick J.~J., 1998, AJ, 115, 1693 (Co98)
\\
De Breuck C., van Breugel W., Minniti D., Miley G., R{\"o}ttgering H., Stanford S.~A., Carilli C., 1999, A\&A, 352, L51 (DeBr99)
\\
De Breuck C., R{\"o}ttgering H., Miley G., van Breugel W., Best P., 2000a, A\&A, 362, 519
\\
De Breuck C., van Breugel W., R{\"o}ttgering H.~J.~A., Miley G., 2000b, A\&AS, 143, 303 (DeBr00)
\\
De Breuck C., et al., 2001, AJ, 121, 1241 (DeBr01)
\\
De Breuck C., van Breugel W., Stanford S.~A., R{\"o}ttgering H., Miley G., Stern D., 2002, AJ, 123, 637 (DeBr02)
\\
De Breuck C., et al., 2004, A\&A, 424, 1 (DeBr04)
\\
De Breuck C., et al., 2010, ApJ, 725, 36 
\\
Douglas J.~N., Bash F.~N., Bozyan F.~A., Torrence G.~W., Wolfe C., 1996, AJ, 111, 1945 (Do96)
\\
Downes A.~J.~B., Peacock J.~A., Savage A., Carrie D.~R., 1986, MNRAS, 218, 31
\\
Downes D., Solomon P. M., Sanders D. B., Evans A. S., 1996, A\&A, 313, 91
\\
Dunlop J.~S., Hughes D.~H., Rawlings S., Eales S.~A., Ward M.~J., 1994, Natur, 370, 347 
\\
Dunlop, J., Peacock, J., Spinrad, H., Dey, A., Jimenez, R., Stern, D., \& Windhorst, R., 1996, nat, 381, 581
\\
Efstathiou A., Rowan-Robinson M., Siebenmorgan R., 2000, MNRAS, 313, 734
\\
Fan X., et al., 2001, AJ, 122, 2833
\\
Fanti C., Fanti R., Ficarra A., Padrielli L., 1974, A\&AS, 18, 147 (Fa74)
\\
Fazio G.~G. et al. 2004, ApJSS, 154, 10
\\
Ficarra A., Grueff G., Tomassetti G., 1985, A\&AS, 59, 255 (Fi85)
\\
Foley R.~J., et al., 2011, arXiv.1101.1286
\\
Gopal-Krishna, Ledoux C., Melnick J., Giraud E., Kulkarni V., Altieri B., 2005, A\&A, 436, 457 (Go05)
\\
Gower J.~F.~R., Scott P.~F., Wills D., 1967, MmRAS, 71, 49 (Go67)
\\
Griffith M., Langston G., Heflin M., Conner S., Lehar J., Burke B., 1990, ApJS, 74, 129 (Gr90)
\\
Gregory P.~C., Condon J.~J., 1991, ApJS, 75, 1011 (Gr91)
\\
Gregory P.~C., Vavasour J.~D., Scott W.~K., Condon J.~J., 1994, ApJS, 90, 173 (Gr94)
\\
Greve T.~R., Stern D., Ivison R.~J., De Breuck C., Kov{\'a}cs A., Bertoldi F., 2007, MNRAS, 382, 48 (Gr07)
\\
Griffith M.~R., Wright A.~E., Burke B.~F., Ekers R.~D., 1994, ApJS, 90, 179 (Gr94)
\\
Griffith M.~R., Wright A.~E., Burke B.~F., Ekers R.~D., 1995, ApJS, 97, 347 (Gr95)
\\
Gutermuth R.~A., Myers P.~C., Megeath S.~T., Allen L.~E., Pipher J.~L., Muzerolle J., Porras A., Winston E., Fazio G., 2008, ApJ, 674, 336
\\
Hales S.~E.~G., Baldwin J.~E., Warner P.~J., 1993, MNRAS, 263, 25 (Ha93)
\\
Hatch N.~A., Overzier R.~A., Kurk J.~D., Miley G.~K., R{\"o}ttgering H.~J.~A., Zirm A.~W., 2009, MNRAS, 395, 114 
\\
Healey S.~E., Romani R.~W., Taylor G.~B., Sadler E.~M., Ricci R., Murphy T., Ulvestad J.~S., Winn J.~N., 2007, ApJS, 171, 61 (He07)
\\
High F.~W. et al., 2010, ApJ, 723, 1736
\\
Hildebrand R.~H., 1983, QJRAS, 24, 267
\\
Humphrey A., Villar-Mart{\'{\i}}n M., Fosbury R., Vernet J., di Serego Alighieri S., 2006, MNRAS, 369, 1103
\\
Humphrey A., Villar-Mart{\'{\i}}n M., Fosbury R., Binette L., Vernet J., De Breuck C., di Serego Alighieri S., 
2007, MNRAS, 375, 705 
\\
Humphrey A., Villar-Mart{\'{\i}}n M., Vernet J., Fosbury R., di Serego Alighieri S., Binette L., 2008, MNRAS, 383, 11 
\\
Inskip K.~J., Best P.~N., R{\"o}ttgering H.~J.~A., Rawlings S., Cotter G., Longair M.~S., 2002, MNRAS, 337, 1407 
\\
Isobe T., Feigelson E.~D., Nelson P.~I., 1986, ApJ, 306, 490
\\
Ivison R.~J., Dunlop J.~S., Smail Ian, Dey Arjun, Liu Michael C., Graham J.~R., 2000, ApJ, 542, 27
\\
Ivison R.~J., et al., 2007, MNRAS, 380, 199 
\\
Ivison R.~J., et al., 2008, MNRAS, 390, 1117 
\\
Jeyakumar S., Saikia D.~J., Pramesh Rao A., Balasubramanian V., 2000, A\&A, 362, 27 (Je00)
\\
Kim S., et al., 2009, ApJ, 695, 809
\\
Knopp G.~P., Chambers K.~C., 1997, ApJS, 109, 367 (Kn97)
\\
Kuehr H., Witzel A., Pauliny-Toth I.~I.~K., Nauber U., 1981, A\&AS, 45, 367 (Ku81)
\\
Lacy M., et al., 2004, ApJS, 154, 166
\\
Large M.~I., Mills B.~Y., Little A.~G., Crawford D.~F., Sutton J.~M., 1981, MNRAS, 194, 693 (La81)
\\
Liu X., Cui L., Luo W.-F., Shi W.-Z., Song H.-G., 2007, A\&A, 470, 97 (Li07)
\\
Mart\'{i}nez-Sansigre A., Lacy M., Sajina A., Rawlings S., 2008, ApJ, 674, 676
\\
Mauch T., Murphy T., Buttery H.~J., Curran J., Hunstead R.~W., Piestrzynski B., Robertson J.~G., Sadler E.~M., 2003, MNRAS, 342, 1117 (Ma03)
\\
McCarthy P.~J., van Breugel W., Spinrad H., Djorgovski S., 1987, ApJ, 321, L29 
\\
McCarthy P.~J., Spinrad H., van Breugel W., 1995, ApJS, 99, 27 
\\
McCarthy P.~J., Kapahi V.~K., van Breugel W., Persson S.~E., Athreya R., Subrahmanya C.~R., 1996, ApJS, 107, 19 (Mc96)
\\
McLure, R.~J., Kukula, M.~J., Dunlop, J.~S., Baum, S.~A., O'Dea, C.~P., \& Hughes, D.~H., 1999, MNRAS, 308, 377 
\\
Menanteau F. et al., 2010, ApJ, 723, 1523
\\
Miley G.~K., Chambers K.~C., van Breugel W.~J.~M., Macchetto F., 1992, ApJ, 401, L69 (Mi92)
\\
Miley G., De Breuck C., 2008, A\&ARv, 15, 67 
\\
Murgia M., Fanti C., Fanti R., Gregorini L., Klein U., Mack K.-H., Vigotti M., 1999, A\&A, 345, 769 
\\
Muxlow T.~W.~B., Pelletier G., Roland J., 1988, A\&A, 206, 237 
\\
Nesvadba N.~P.~H., et al., 2006, ApJ, 650, 661
\\
Nesvadba N.~P.~H., et al., 2009, MNRAS, 395, L16 
\\
Overzier R.~A., R{\"o}ttgering H.~J.~A., Kurk J.~D., De Breuck C., 2001, A\&A, 367, L5 (Ov01)
\\
Pentericci L., Roettgering H.~J.~A., Miley G.~K., Carilli C.~L., McCarthy P., 1997, A\&A, 326, 580 (Pe97)
\\
Pentericci L., Van Reeven W., Carilli C.~L., R{\"o}ttgering H.~J.~A., Miley G.~K., 2000, A\&AS, 145, 121 (Pe00)
\\
Pentericci L., McCarthy P.~J., R{\"o}ttgering H.~J.~A., Miley G.~K., van Breugel W.~J.~M., Fosbury R., 2001, ApJS, 135, 63
\\
Pilkington J.~D.~H., Scott P.~F., 1965, MmRAS, 69, 183 (Pi65)
\\
Priddey R.~S., Isaak K.~G., McMahon R.~G., Robson E.~I., Pearson C.~P., 2003, MNRAS, 344, L74 
\\
Priddey R.~S., Ivison R.~J., Isaak K.~G., 2008, MNRAS, 383, 289 
\\
Reuland M., et al., 2003a, ApJ, 592, 755 (Re03)
\\
Reuland Michiel, van Breugel Wil, R\"ottgering Huub, de Vries Wim, De Breuck Carlos, Stern Daniel, 2003b, ApJ, 582L, 71
\\
Reuland M., R{\"o}ttgering H., van Breugel W., De Breuck C., 2004, MNRAS, 353, 377 (Re04)
\\
Reuland Michiel, 2005, PhD Thesis, Leiden Observatory, Leiden University
\\
Rickett B.~J., Lazio T.~J.~W., Ghigo F.~D., 2006, ApJS, 165, 439 (Ri06a)
\\
Ricci R., Prandoni I., Gruppioni C., Sault R.~J., de Zotti G., 2006, A\&A, 445, 465 (Ri06b)
\\
Rieke G.~H. et al., 2004, ApJSS, 154, 25
\\
R\"ottgering, H.~J.~A., van Ojik, R., Miley, G.~K., Chambers, K.~C., van Breugel, W.~J.~M., \& de Koff, S., 1997, A\&A, 326, 505 (R\"o97)
\\
Scott K.~S., et al., 2010, MNRAS, 405, 2260 
\\
Seymour N., et al., 2007, ApJS, 171, 353
\\
Slee O.~B., 1995, AuJPh, 48, 143 (Sl95)
\\
Sol{\'o}rzano-I{\~n}arrea C., Best P.~N., R{\"o}ttgering H.~J.~A., Cimatti A., 2004, MNRAS, 351, 997 
\\
Stanghellini C., O'Dea C.~P., Dallacasa D., Baum S.~A., Fanti R., Fanti C., 1998, A\&AS, 131, 303 (St98)
\\
Stanghellini C., O'Dea C.~P., Dallacasa D., Cassaro P., Baum S.~A., Fanti R., Fanti C., 2005, A\&A, 443, 891 (St05)
\\
Stern D., et al., 2005, ApJ, 631, 163
\\
Stevens J.~A., et al., 2003, Natur, 425, 264 
\\
Stevens J.~A., Jarvis M.~J., Coppin K.~E.~K., Page M.~J., Greve T.~R., Carrera F.~J., Ivison R.~J., 2010, MNRAS, 405, 2623 
\\
Stiavelli M., et al., 2005, ApJ, 622, L1 
\\
van Breugel W., Miley G., Heckman T., Butcher H., Bridle A., 1985, ApJ, 290, 496 
\\
van Breugel W., De Breuck C., Stanford S.~A., Stern D., R{\"o}ttgering H., Miley G., 1999, ApJ, 518, L61 (vB99)
\\
van Ojik R., Roettgering H.~J.~A., Miley G.~K., Hunstead R.~W., 1997, A\&A, 317, 358 
\\
Venemans B.~P., et al., 2007, A\&A, 461, 823 (Ve07)
\\
Vernet J., Fosbury R.~A.~E., Villar-Mart{\'{\i}}n M., Cohen M.~H., Cimatti A., di Serego Alighieri S., Goodrich R.~W., 2001, A\&A, 366, 7 
\\
Vieira J.~D., et al., 2010, ApJ, 719, 763 
\\
Villar-Mart{\'{\i}}n M., Fosbury R.~A.~E., Binette L., Tadhunter C.~N., Rocca-Volmerange B., 1999, A\&A, 351, 47 (Vi99)
\\
Villar-Mart{\'{\i}}n M., Vernet J., di Serego Alighieri S., Fosbury R., Humphrey A., Pentericci L., 2003, MNRAS, 346, 273 
\\
Villar-Mart{\'{\i}}n M., et al., 2006, MNRAS, 366, L1
\\
Villar-Mart{\'{\i}}n M., Humphrey A., De Breuck C., Fosbury R., Binette L., Vernet J., 2007, MNRAS, 375, 1299
\\
Viner M.~R., Erickson W.~C., 1975, AJ, 80, 931 (Vi75)
\\
Werner M.~W. et al., 2004, ApJSS, 154, 1
\\
White R.~L., Becker R.~H., 1992, ApJS, 79, 331 (Wh92)
\\
Willott C.~J., Rawlings S., Archibald E.~N., Dunlop J.~S., 2002, MNRAS, 331, 435
\\
Willott C.~J., Delfosse X., Forveille T., Delorme P., Gwyn S.~D.~J., 2005, ApJ, 633, 630 
\\
Wilson G.~W., et al., 2008, MNRAS, 386, 807 
\\
Wright A., Otrupcek R., 1990, Parkes Catalogue, Australia Telescope National Facility (Wr90)
\\
Wright A.~E., Griffith M.~R., Burke B.~F., Ekers R.~D., 1994, ApJS, 91, 111 (Wr94)
\\
Xiang L., Dallacasa D., Cassaro P., Jiang D., Reynolds C., 2005, A\&A, 434, 123 (Xi05)
\\
Xiang L., Reynolds C., Strom R.~G., Dallacasa D., 2006, A\&A, 454, 729 (Xi06)
\\
Yun M.~S., et al., 2008, MNRAS, 389,333
\\
Yun M.~S., et al., 2011, MNRAS, submitted

\label{lastpage}

\end{document}